\documentclass[twocolumn]{aastex631}

\usepackage{romannum}
\usepackage{soul}

\received{May 4, 2025}
\revised{Oct 2, 2025}
\submitjournal{AAS Journals}

\shorttitle{Comprehensive Analysis of WASP-96}
\shortauthors{Wang et al.}
\usepackage{amsmath}
\usepackage{multirow}
\graphicspath{{./}{mainFigures/}}

\begin{document}

\title{A Comprehensive Analysis of the Panchromatic Transmission Spectrum of the Hot-Saturn WASP-96 b: Nondetection of Haze, Possible Sodium Limb Asymmetry, Stellar Characterization, and Formation History}

\author[0000-0002-6379-3816]{Le-Chris Wang}
\affiliation{Department of Astrophysical Sciences, Princeton University, 4 Ivy Lane, Princeton, NJ 08544, USA}
\affiliation{William H. Miller III Department of Physics \& Astronomy, Johns Hopkins University, 3400 N Charles St, Baltimore, MD 21218, USA}
\email{lechris.wang@princeton.edu}

\author[0000-0003-4408-0463]{Zafar Rustamkulov}
\affiliation{Department of Earth and Planetary Science, Johns Hopkins University, 3400 N. Charles Street, Baltimore, MD 21218, USA}

\author[0000-0001-6050-7645]{David K. Sing}
\affiliation{William H. Miller III Department of Physics \& Astronomy, Johns Hopkins University, 3400 N Charles St, Baltimore, MD 21218, USA}
\affiliation{Department of Earth and Planetary Science, Johns Hopkins University, 3400 N. Charles Street, Baltimore, MD 21218, USA}

\author[0000-0003-3667-8633]{Joshua Lothringer}
\affiliation{William H. Miller III Department of Physics \& Astronomy, Johns Hopkins University, 3400 N Charles St, Baltimore, MD 21218, USA}
\affiliation{Space Telescope Science Institute, 3700 San Martin Drive, Baltimore, MD 21218, USA}

\author[0000-0003-0473-6931]{Patrick McCreery}
\affiliation{William H. Miller III Department of Physics \& Astronomy, Johns Hopkins University, 3400 N Charles St, Baltimore, MD 21218, USA}

\author[0000-0002-5113-8558]{Daniel Thorngren}
\affiliation{William H. Miller III Department of Physics \& Astronomy, Johns Hopkins University, 3400 N Charles St, Baltimore, MD 21218, USA}

\author[0000-0003-4157-832X]{Munazza K. Alam}
\affiliation{Space Telescope Science Institute, 3700 San Martin Drive, Baltimore, MD 21218, USA}



\begin{abstract}
\noindent We conduct a reanalysis of the JWST NIRISS/SOSS observation of the hot-Saturn WASP-96 b. Initial analysis of this data revealed an enhanced Rayleigh scattering slope at the blue end of the transmission spectrum, suggesting the presence of hazes at high altitudes. In this work, we report non-detection of this slope, confirming an atmosphere clear of high-altitude aerosols consistent with the pre-JWST results. Also contrary to the initial result, our results indicate the presence of gray cloud deck, although at relatively low altitudes/high pressures. We further combined the NIRISS/SOSS spectrum with VLT, HST, and Spitzer to produce a transmission spectrum from 0.35 $\mu$m to 5 $\mu$m. We constrain the mass fraction of multiple chemical species, including: H$_2$O $=-2.62^{+0.43}_{-0.42}$, K $=-5.76^{+1.05}_{-1.13}$, and Na $=-3.40^{+0.90}_{-0.92}$. C/O ratio and metallicity are tentatively constrained at substellar values ($\text{C/O}_{planet}=0.57^{+0.07}_{-0.12}$ and $\text{[Fe/H]}_{planet}=0.01^{+0.46}_{-0.52}$ compared to $\text{C/O}_{star} = 0.92\pm0.25$ and $\text{[Fe/H]}_{star}=0.24\pm0.05$). Inputing these composition constraints to interior structure models, we constrain a core mass of $43^{+8}_{-15}$ M$_\oplus$. This, in addition to our inferred super-stellar refractory-to-oxygen ratio ($\Delta\log_{10}(R/O)=1.48^{+0.57}_{-0.62}$) and substellar C/O ratio, suggests that the core of WASP-96 b likely formed outside of water iceline, underwent disk-driven migration, and accreted its atmosphere inside the carbon soot line. We find evidence of atmospheric leading-trailing terminator asymmetries in the broadened sodium absorption feature with a transit time offset of 50 seconds, while the water features appear symmetric. CH$_4$, CO, and CO$_2$ remain unconstrained due to spectral coverage limits. Upcoming JWST NIRSpec/G395H observations (ID 4082, PI: M. Radica) will be crucial for constraining these carbon-bearing species.
\end{abstract}


\section{Introduction} \label{sec:intro}
The combination of ground-based and space-based observations has contributed significantly to the study of exoplanet atmospheres. State-of-the-art space-based telescopes, such as the Hubble Space Telescope (HST) and the now-decommissioned Spitzer, have detected a rich diversity of molecular and atomic species through high-precision photometry and spectroscopy (e.g. \citealt{charbonneau2002detection,2015MNRAS.446.2428S,sing2016,garhart2020,2022MNRAS.515.3037N,2023MNRAS.522.5062B}). HST spans a broad wavelength range from the ultraviolet to the near-infrared, with instruments such as Cosmic Origins Spectrograph (COS) ($0.08 - 0.32$ $\mu m$) and Space Telescope Imaging Spectrograph (STIS) ($0.12 - 1.03 ~\mu m$) enabling access to far-UV features like Lyman-alpha (e.g. \citealt{ehrenreich2008,bourrier2018}) and refractory metal lines (e.g. \citealt{lothringer2020,gressier2023,lothringer2025}), and Wide Field Camera 3 (WFC3) providing low-resolution optical to near-infrared spectroscopy between 0.2 and 1.7 $\mu m$. Spitzer, during its cryogenic phase, provided both photometry and spectroscopy across $3 – 40$ $\mu m$, including low-resolution spectra for a small number of exoplanets \citep{grillmair2007, swain2008}. Following its coolant loss in 2009, Spitzer operated in its “warm” mission phase using the 3.6 and 4.5 $\mu m$ channels, which still contributed significantly to exoplanet phase curves and transit observations until its decommissioning in 2020. In the meantime, ground-based telescopes have provided complementary data that advance our understanding of exoplanet atmospheres (e.g. \citealt{barman2015simultaneous,2018Natur.557..526N,ehrenreich2020nightside,2022AJ....164..134M,2024arXiv241018663P}), offering key observational advantages such as high-resolution spectroscopy and cost-effectiveness.


JWST is revolutionizing the field of exoplanet atmospheres by extending the spectral window to the infrared (0.6 $\mu m$ to 14 $\mu m$) at moderate resolution (R $\sim$ 50-2600, compared to R $\sim$ 20-100 for HST). In its three years of operations, JWST has identified new molecular features that have never been constrained before, such as SO$_2$, CO$_2$, and CH$_4$ (e.g. \citealt{jwsters2023,ahrer2023,2023Natur.614..659R,alderson2023,feinstein2023,sing2024,benneke2024,schmidt2025, kirk2025}). The Near Infrared Imager and Slitless Spectrograph Single Object Slitless Spectroscopy (NIRISS/SOSS) mode \citep{Albert_2023} is one of the workhorse modes on JWST for transmission spectroscopy. Covering 0.6 $\mu m$ to 2.85 $\mu m$ at moderate resolution (R $\sim$ 700), NIRISS/SOSS is able to constrain multiple water bands (e.g. \citealt{fu2022,pg2024,louie2025}), methane (e.g. \citealt{schmidt2025}), and tentatively CO and CO$_2$ (e.g. \citealt{feinstein2023,2023MNRAS.524..817T}). By extending wavelength coverage to the optical, SOSS also covers alkali metals such as K (e.g. \citealt{feinstein2023,2023MNRAS.524..817T,ft2025}) and Li and is able to reveal the presence of clouds and hazes (e.g. \citealt{fu2022,coulombe2025}). However, since the wavelength coverage of all JWST instruments stops at 0.6 $\mu m$, JWST is only able to detect the red wing of the Na feature. Thus, combining JWST with other instruments that probe bluer wavelengths, such as the Very Large Telescope (VLT), is necessary to detect Na confidently (e.g. \citealt{feinstein2023, fisher2024}). Historically, the combination of HST transmission spectra with Spitzer Infrared Array Photometry (IRAC) has played a key role in constraining the carbon-to-oxygen (C/O) ratio in exoplanet atmospheres (e.g. \citealt{spake2021,2022MNRAS.515.3037N}), as the Spitzer points at the 3.6 and 4.5 $\mu m$ channels overlap strongly with the most prominent absorption regions of CO, CO$_2$, and CH$_4$. Therefore, in cases where only NIRISS/SOSS data (0.6–2.8 $\mu m$) are available, combining the SOSS spectrum with archival Spitzer photometric data could similarly improve constraints on the C/O ratio, particularly by providing sensitivity to CO and CO$_2$ features that fall outside the NIRISS bandpass.


In this paper, we present a re-analysis of the NIRISS/SOSS transit spectrum of WASP-96 b. WASP-96 b \citep{discovery2014} is a hot-Saturn with a mass of 0.48 $\pm$ 0.03 M$_{\text{J}}$ and a radius of 1.20 $\pm$ 0.06 R$_{\text{J}}$ orbiting around a G-type star in 3.42 days. The low density and relatively high equilibrium (1285 $\pm$ 40 K, \citealt{Hellier2014}) temperature of WASP-96 b make it an excellent candidate for transmission spectroscopy. Early observations with FOcal Reducer/low dispersion Spectrograph 2 (FORS2) on VLT revealed a clear pressure-broadened sodium feature, suggesting WASP-96 b is mostly free of clouds and hazes \citep{2018Natur.557..526N}. The measured sodium abundance is consistent with the stellar value. Later, \cite{2021AJ....161....4Y} and \cite{2022MNRAS.515.3037N} independently analyzed the transmission spectrum of WASP-96 b with HST/WFC3 and Spitzer/IRAC. These authors found solar-to-super-solar sodium and oxygen abundance, suggesting the bulk metallicity of WASP-96 b is consistent with the mass-metallicity trend in the solar system \citep{thorngren2016}. \cite{2022MNRAS.515.3037N} further found a sub-solar C/O assuming chemical equilibrium. Moreover, \cite{2022AJ....164..134M} independently confirmed the pressure-broadened sodium feature and the solar-to-super-solar sodium and water abundance using Magellan/IMACS.

As a part of the JWST Early Release Observations (ERO) program \citep{2022ApJ...936L..14P}, WASP-96 b was observed by the JWST NIRISS/SOSS mode, extending the wavelenth coverage from HST's 1.7 $\mu m$ to 2.8$\mu m$. \cite{2023MNRAS.524..835R} and \cite{2023MNRAS.524..817T} presented the first analysis of the JWST NIRISS/SOSS data. These authors constrained potassium, solar-to-super-solar metallicity, and a solar C/O in WASP-96 b's atmosphere. However, contrary to pre-JWST results, these authors found an enhanced blueward slope in the SOSS spectrum. \cite{2023MNRAS.524..817T} attributed the slope to the enhanced Rayleigh scattering and concluded that small, inhomogeneous aerosols could be present at high altitudes, while \cite{2023MNRAS.524..835R} suggested that the slope could also be caused by the red end of the sodium wing.

Given the abundance of high-quality pre-JWST observations, combining legacy datasets with new JWST spectra is particularly attractive for analyzing WASP-96 b’s atmosphere. The comprehensive optical-to-near-infrared coverage enabled by VLT, HST, and Spitzer provides constraints on alkali metals, water, and C/O ratio. The addition of JWST/NIRISS SOSS data probes the infrared region with higher precision and resolution. In this work, we reanalyze the NIRISS/SOSS data of WASP-96 b and combine the resulting transmission spectrum with VLT, HST, and Spitzer data. The Magellan/IMACS data is not included in the transmission spectrum analysis because it overlaps the VLT wavelengths (0.48 - 0.83 $\mu m$ for Magellan and 0.36 - 0.82 $\mu m$ for VLT), and the reduced spectrum from \cite{2022AJ....164..134M} is consistent with that from VLT. We analyze the consistency between NIRISS/SOSS data and previous legacy datasets and comprehensively analyze the atmospheric properties of WASP-96 b from the resulting panchromatic transmission spectrum from 0.35 - 5 $\mu m$. We describe the observations and data reduction in Section \ref{sec:obs}. We describe the processes to fit the light curves in Section \ref{sec:lc}. We present the transmission spectrum and describe the atmospheric analysis methods in Section \ref{sec:analysis}. We show and analyze the results in Section \ref{sec:result}. We present a refined host star characterization analysis in Section \ref{sec:stellar}. Finally, we discuss and summarize our findings in Section \ref{sec:discussion} and \ref{sec:conclusion}.

\section{Observations and Data Reductions}\label{sec:obs}

\subsection{Observations}

\subsubsection{JWST NIRISS SOSS}
An observation of WASP-96 b with NIRISS SOSS was made as a part of the JWST Early Release Observations \citep{2022ApJ...936L..14P}. The 6.4-hour Time Series Observation (TSO) started on UTC June 21, 2022, and includes 280 NISRAPID integrations with 14 groups per integration. A standard GR700XD/CLEAR combination was used together with the SUBSTRIP256 subarray to capture diffraction of all three spectral orders (covering $0.6-2.8~\mu m$). The observation also includes an optional second exposure with the GR700XD/F277W combination that only permits spectral trace with wavelengths higher than $2.6 ~\mu m$. This second exposure included 11 integrations and lasted for about 0.25 hours. The specific observations analyzed can be accessed via \dataset[doi: 10.17909/xpp6-xh70]{https://doi.org/10.17909/xpp6-xh70}.

\subsubsection{VLT $+$ HST WFC3 $+$ Spitzer IRAC}

Transit observations of WASP-96 b with the 8.2-meter VLT/FORS spectrograph were obtained on UTC July 29th and August 22nd, 2017 as a part of Large Program 199.C-0467 (PI: Nikolov, \citealt{2018Natur.557..526N}). The observation uses the multi-object-spectroscopy mode with grisms 600B and 600RI on the first and second nights respectively, covering $0.36-0.82~\mu m$.
 
Two primary transit observations of WASP-96 b with WFC3 on the \textit{Hubble Space Telescope (HST)} were made as a part of Program 15469 (PI: Nikolov, \citealt{nikolovhst}). The first observation occurred on UTC December 18th, 2018 with dispersive element G102 ($ 0.8 - 1.1 ~\mu m$). The second observation occurred on UTC December 28th, 2018 with dispersive element G141 ($1.1 - 1.7 ~ \mu m$). Together, both observations cover the spectral range of $0.8 - 1.7 ~\mu m$. The specific descriptions of the observation are detailed in \cite{2022MNRAS.515.3037N}.

Two primary transits of WASP-96 b with \textit{Spitzer Space Telescope} Infrared Array Camera (IRAC) in the 3.6 and 4.5 $\mu m$ channels were observed separately during two transit events as part of Program 14255 (PI: Nikolov, \citealt{nikolovspitzer}). The two transit observations started on UTC October 22nd, 2019, and October 29th, 2019, respectively.

The data for the transmission spectra from observations with the three observatories described above were obtained both through private communication with Nikolov and from \cite{2018Natur.557..526N} and \cite{2022MNRAS.515.3037N}.

\subsection{NIRISS SOSS Data Reduction}
We use the Fast InfraRed Exoplanet Fitting for Lightcurves (FIREFLy) pipeline \citep{2022ApJ...928L...7R, 2023Natur.614..659R} for our reduction. FIREFLy was originally developed and optimized for JWST NIRSpec PRISM/G395H observations, and we have updated it to be capable of reducing NIRISS SOSS data with almost identical workflow \citep{liuwang2025}. We follow almost the same procedures described in \cite{liuwang2025}, which we describe briefly below.

First, we perform detector-level corrections to the uncalibrated data, where we follow most of the calibration steps as the \verb|jwst| pipeline (version 1.14.0) 
using default values. Most notably, superbias subtraction, group-level column-correlated $1/f$ noise subtraction with the temporal median method, and ramp fitting are performed. With the exception of group-level $1/f$ subtraction, all the steps follow from the STScI pipeline \citep{2023zndo...6984365B}. For the temporal median method, we take the median of 7 consecutive integrations for each group, treating this as a reference "uncontaminated" image. Subtracting each group image from this median image isolates the column-correlated 1/f noise component, which is then removed from the original group data. We find group-level $1/f$ subtraction typically lowers the noise by $\sim$5\%.

We then proceed to stage 2 of FIREFLy, where we perform bad pixel cleaning, background removal, and 1D extraction. The shape of the SOSS sky background is formed when zodiacal light illuminates the pick-off mirror (POM) and is then dispersed by the GR700XD grism. A sharp, step-like feature appears within this background at 2.1 µm, which is caused by the physical edge of the POM. To remove this background structure, we use the background model publicly available on \verb|jwst-docs|\footnote{\href{https://jwst-docs.stsci.edu/}{https://jwst-docs.stsci.edu/}}. We mask out the spectral traces and order contaminants and fit the background model to the flux level at $y\in[200, 250]$ on either side of the flux jump ($x\simeq 700$). After the background is removed, we fully destrip the $1/f$ noise at the integration level by using both a temporal median method and a PSF mask. The PSF mask is constructed by applying a 5th percentile flux threshold, carefully chosen so that both the stellar trace and the contamination from the field stars are included; then, column median values for pixels outside of the PSF are subtracted. The PSF mask method also helps remove any remaining background. We use cross-correlation to measure the positional shift of the stellar trace and shift stabilize the images with flux-conserving cubic spline interpolations. We detect only minor, random shifts (on the order 0.1 mas) along both axes. The measured positional shifts along both axes are incorporated to the systematics model in the light curve fitting step (see section \ref{sec:lc}).




Since SOSS is a slitless spectrograph, starlight in the background within the field of view will unavoidably be dispersed onto the detector. If the dispersed starlight overlaps with the spectral trace of the target, it will introduce contamination to the transmission spectrum. This is the case for the WASP-96 b observation. As shown in Figure 3 of \cite{2023MNRAS.524..835R}, a first-order contaminant lying below the order 1 spectral trace for roughly $x\in[0, 1000]$ crosses the order 1 spectral trace for $x\gtrsim 1000$. This will introduce a wavelength-dependent dilution to the transmission spectra, primarily affecting pixels between $1.3$ and $ 1.5$ $\mu m$. Moreover, a zeroth-order contaminant centered at $x\approx790$ also overlaps with the order 1 spectral trace partially (corresponding to wavelengths at $\sim2.04~\mu m$), which will also introduce dilution in the transmission spectra at wavelengths corresponding to those affected pixels.

\begin{figure*}[t]
    \centering
    \plotone{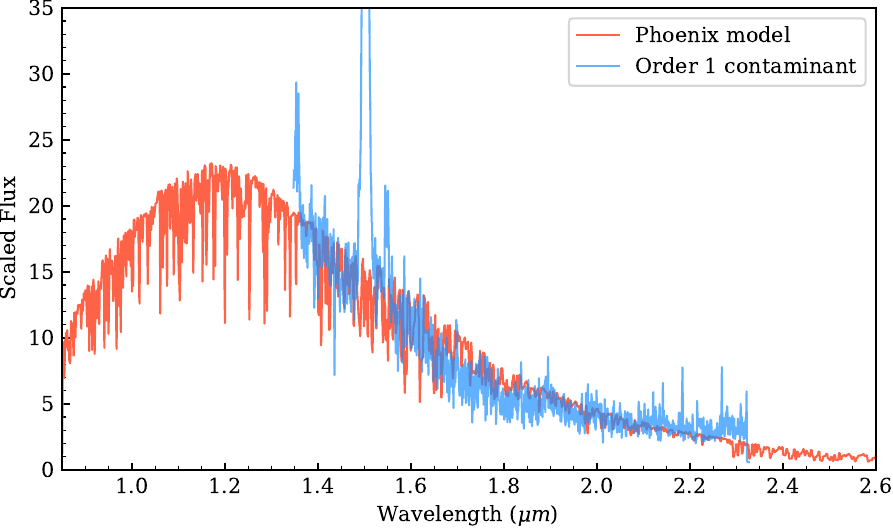}
    \caption{Comparison between the extracted order 1 contaminant spectrum and the best fitting PHOENIX stellar model spectrum corrected for SOSS throughput. The PHOENIX stellar model has $T_{eff} = 3900~K$, log g $= 4.5$, and [Fe/H] = 0.5.  The large spike at $\sim$ 1.5 $\mu m$ is due to a zeroth order contaminant overlapping with the first order contaminant.}
    \label{fig:phoenix}
\end{figure*}

In order to correct those contaminants, we primarily follow the post-processing steps described in \cite{2023MNRAS.524..835R}. We identify that the zeroth-order contaminant spans about 11 columns in the horizontal direction. To correct the zeroth-order contaminants, we stack 10 columns on either side of the contaminated regions and linearly interpolate over the contaminated region to roughly estimate the counts if the region were uncontaminated. To correct the first-order contaminants, we subtract the target order 1 trace modeled by the \verb|ATOCA| algorithm (embedded in the \verb|extract1d| method in the STScI pipeline \citealt{2023zndo...6984365B}), which allows the first-order contaminant to be better revealed. We manually fit the first-order target centroid to the order 1 contaminant, which gives us the wavelength solution. We find offset values identical to those reported by \cite{2023MNRAS.524..835R}. Finally, we fit PHOENIX stellar models \citep{2013A&A...553A...6H} corrected for the instrument throughput to the extracted order 1 contaminant spectrum, and find the model with parameters $T_{eff} = 3900~K, log~g = 4.5$, and [Fe/H] $= 0.5$ a good fit (shown in Figure \ref{fig:phoenix}), consistent with the parameters of the potential contaminant star in the field and the result reported by \cite{2023MNRAS.524..835R}. We smooth the model PHOENIX spectrum and use it to approximate a model order 1 contaminant trace, which is then subtracted.

\begin{figure*}[t]
    \centering
    \plotone{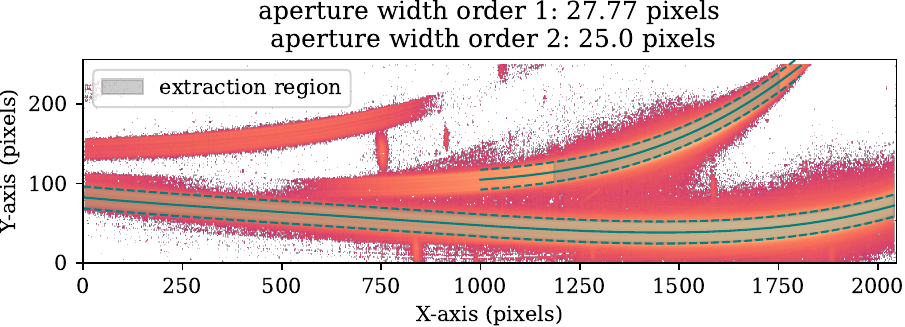}
    \caption{Extraction of 1D spectrum for order 1 and order 2 spectral traces with simple box extraction. The box widths are optimized so that the scatter for the out-of-transit light curves is minimized.}
    \label{fig:box}
\end{figure*}

Finally, we use the box extraction to extract spectrophotometry for order 1 and order 2 spectral traces respectively, with the width of the extraction boxes optimized to minimize the scatter of off-transit data points. The optimally determined box width for order 1 is $\sim$ 28 pixels while that for order 2 is 25 pixels. We do not correct for the contamination from the order overlap at the left side of the detector, as it only introduces negligible contamination to the transmission spectrum \citep{2022PASP..134i4502D, 2022PASP..134j4502R}.  For order 2 spectral trace, we only extract wavelengths $\in [0.6, 0.85]~\mu m$. The extraction box for both orders is shown in Figure \ref{fig:box}.

\section{Light Curve Analysis}\label{sec:lc}
\begin{figure*}
    \centering    \plotone{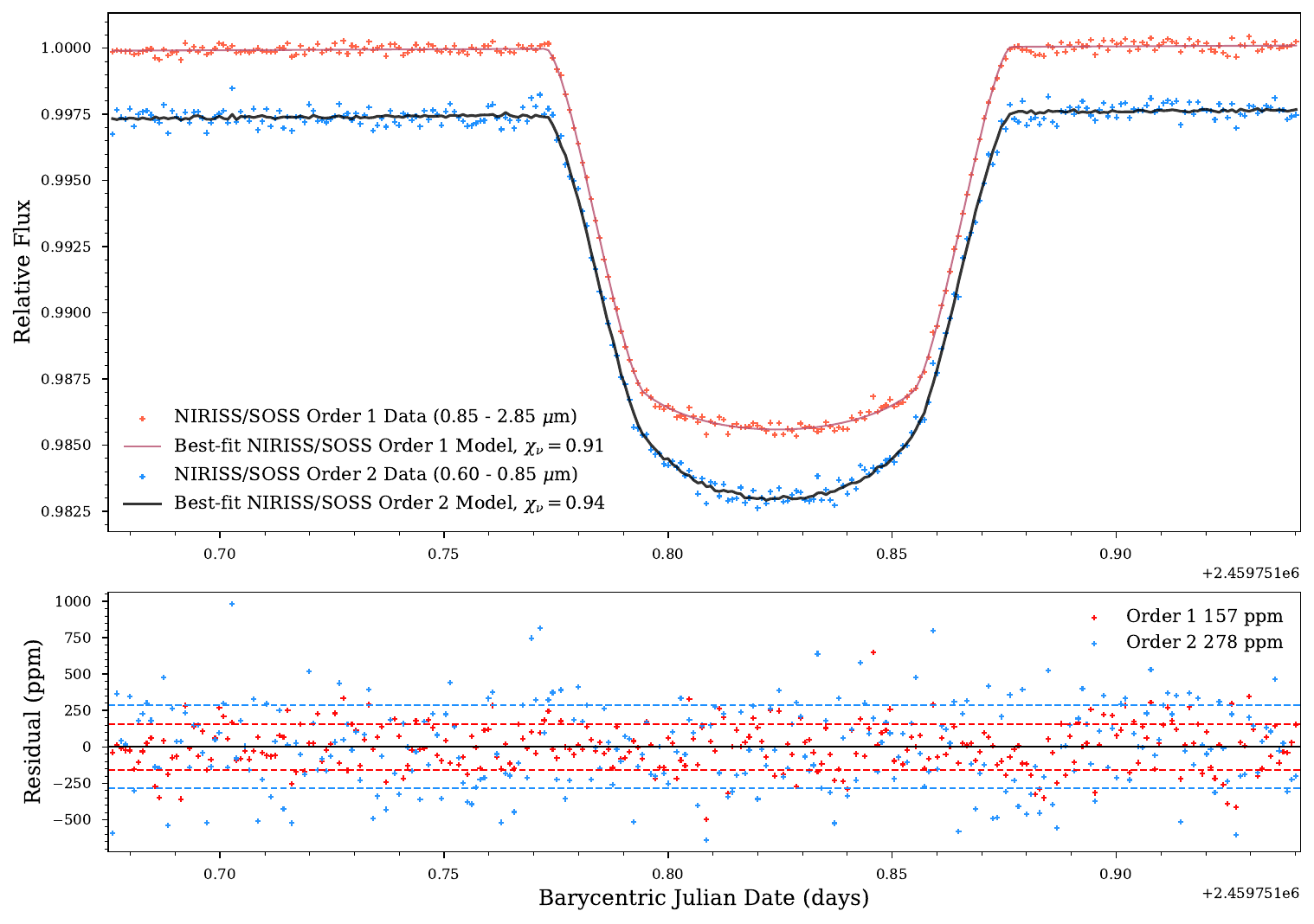}
    \caption{Best fitting white light curves with residuals. \textit{Top}: the best-fit white light curves for SOSS order 1 and order 2 overplotted with the data. The reduced chi-square for order 1 is 0.91 and that for order 2 is 0.94. \textit{Bottom}: The residuals for the white light curve fits. The scatter for order 1 is 157 ppm, and that for order 2 is 278 ppm. The red and blue dashed lines show median scatter of order 1 and order 2 residuals, respectively.}
    \label{fig:wlc}
\end{figure*}

We sum the flux in the extraction region in Figure \ref{fig:box}  to obtain the white light curve (Figure \ref{fig:wlc}) for each order by inspection. We trim the data in both time and wavelength to exclude some bad data points. The first 2 integrations are excluded because of the non-linear baseline flux trend they show. For the first-order spectral trace, we exclude the first five and last five columns. We use the out-of-transit part of the light curve to determine the best systematics models for detrending by running linear regression with different systematics combinations. We select the best model based on the Bayesian Information Criterion (BIC) \citep{kass1995bayes}. For order 1 white light curve, the best systematics returned is a first-order linear term in time, $lin_1$. For order 2, the best systematics is a combination of X-axis trace shift, $xs$, and $lin_1$.

We fit the white light curve with \verb|batman| \citep{2015PASP..127.1161K}. We fix the orbital period \textit{p} and eccentricity \textit{e} to 3.4252602 days and 0 following \cite{2014MNRAS.440.1982H}. We set the scaled semi-major axis $a/R_*$, impact parameter $b$, and the mid-transit time $t_0$ as free parameters, and jointly fit them with the other parameters: base flux $f_0$, scaled planetary radius $(R_p/R_*)^2$, the best systematics models, and the quadratic limb darkening coefficients $q_1$ and $q_2$ following the parametrization from \cite{2013MNRAS.435.2152K}:
\begin{equation}
    q_1 = (u_1+u_2)^2 \quad q_2 = \frac{u_1}{2(u_1+u_2)}
    \label{ldc}
\end{equation}

\begin{figure*}
    \centering
    \plotone{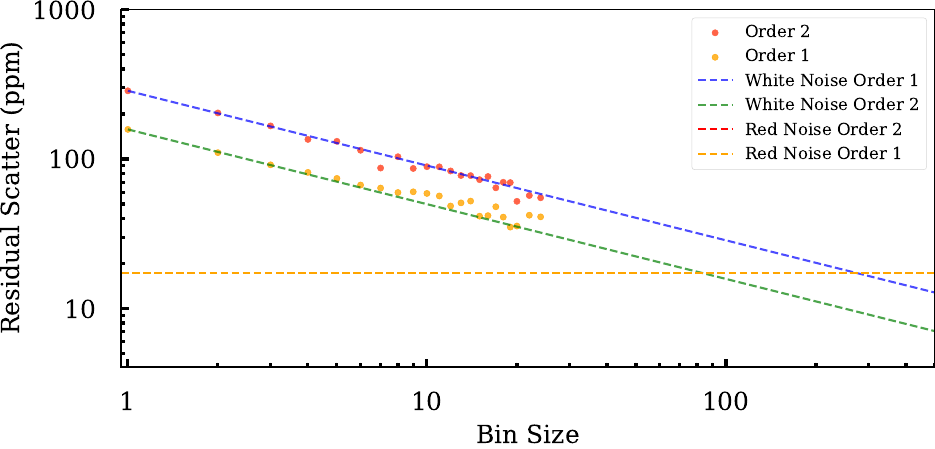}
    \caption{The relationship between bin size and residual scatter for white light curve of order 1 and the white light curve of order 2. The blue and green dashed lines are the theoretical decrease for white noise (proportional to $1/\sqrt{N}$) for order 1 and order 2, respectively, and the horizontal dashed lines are estimates of the red noise. The maximum bin size is chosen such that there are at least 10 data points to measure the standard deviations reliably. The residuals generally trace the photon noise. The red noise level for order 2 is $\sim$ 0 and is therefore off the scale of this plot.}
    \label{fig:residuals}
\end{figure*}

The limb darkening coefficients are approximated using the 3D stellar model from \cite{2015A&A...573A..90M} before fitting. We perform the joint fit by first using the Levenberg-Marquart least-squares minimization algorithm for an initial fit, and the Markov Chain Monte Carlo (MCMC) with wide uninformative priors to fully explore the parameter space. Both algorithms are provided by the \verb|lmfit| package \citep{2014zndo.....11813N}. The fitted white light curves for both orders are shown in Figure \ref{fig:wlc}. The best-fit parameters are summarized in Table \ref{tab:wlc}. The values from our fit are statistically consistent with those reported by \cite{2023MNRAS.524..835R} within 1 $\sigma$. We also find the residuals trace pure photon noise as they are binned down, as shown in Figure \ref{fig:residuals}.

To extract the transmission spectrum, we bin the spectrophotometry in wavelength by 10 pixels and fit the light curve at each bin individually. We fix $t_0$, $a/R_*$, and $b$ to the fitted median values from the white light curve fit for each order separately. We find the model limb darkening coefficients theoretically derived from the 3D stellar model \citep{2015A&A...573A..90M} trace the fitted limb darkening coefficients well, so we fix the limb darkening coefficients to the model values. We only set $(R_p/R_*)^2$, the systematics model (same as that from the white light curve fit for each order), and $f_0$ as free parameters. We find least squares method suffice to fit the light curves at each wavelength bins, as the posterior distribution is well-approximated by a multivariate Gaussian.

\begin{table*}
\caption{Best-fitting white light curve parameters}
\label{tab:wlc}
    \begin{tabular}{ccccccc}
        \hline
        \hline
        Order & $\rm t_0~[BJD - 2459751] $ & $\rm (R_p/R_*)^2$ & $\rm b$ & $\rm a/R_*$ & $q_1$ & $q_2$\\
        \hline
        1 & $0.82468 \pm 4\times10^{-5}$ & $0.0142\pm2\times10^{-4}$ & $0.7265\pm0.0047$ & $8.994\pm0.047$ & $0.17 \pm 0.04$& $0.27\pm 0.22$\\
        2 & $0.82471 \pm 7\times 10^{-5}$ & $0.0141\pm3\times10^{-4}$ & $0.7186\pm0.0108$ & $8.961\pm 0.103$ & $0.50\pm 0.13$& $0.25\pm 0.18$ \\
        \hline
    \end{tabular}
\end{table*}

\section{Spectral Analysis}\label{sec:analysis}

\subsection{Observed Transmission Spectrum}

\begin{figure*}[t]
    \centering
    \plotone{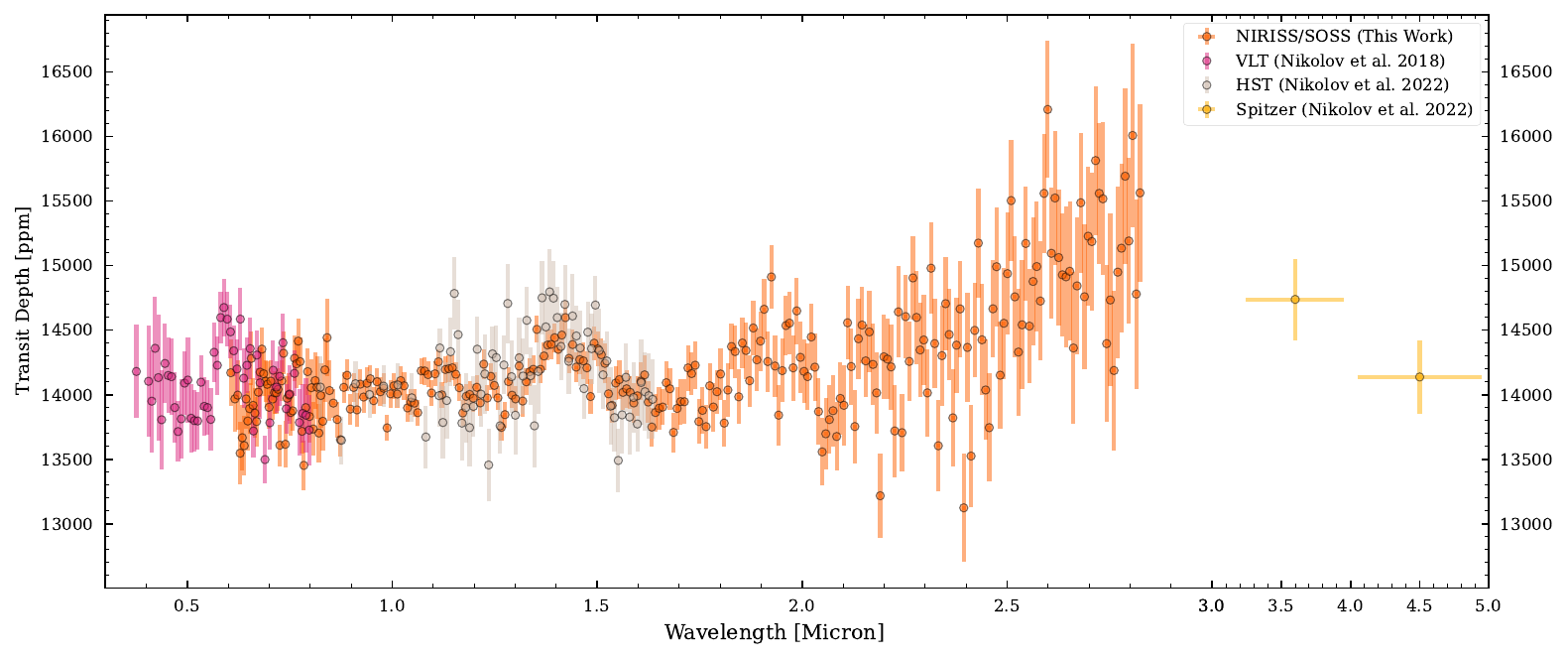}
    \caption{SOSS transmission spectrum extracted with FIREFLy combined with VLT \citep{2018Natur.557..526N}, Hubble \citep{2022MNRAS.515.3037N}, and Spitzer \citep{2022MNRAS.515.3037N} spectrum. At overlapping wavelengths, the VLT and Hubble spectra are consistent with the SOSS spectrum.}
    \label{fig:transmission spectrum}
\end{figure*}

We combine the transmission spectrum extracted from SOSS with that produced from HST, VLT, and Sptizer \citep{2018Natur.557..526N, 2022MNRAS.515.3037N}, as shown in Figure \ref{fig:transmission spectrum}. We find the transmission spectrum from SOSS traces well the spectrum from VLT and Hubble at overlapping wavelengths. The spectrum from SOSS provides much higher resolution and smaller error bars, which could better constrain the abundances of possible atmospheric species. We do not find any offsets fitted between the VLT/Hubble spectrum and the SOSS spectrum.

\begin{figure*}[t]
    \centering
    \plotone{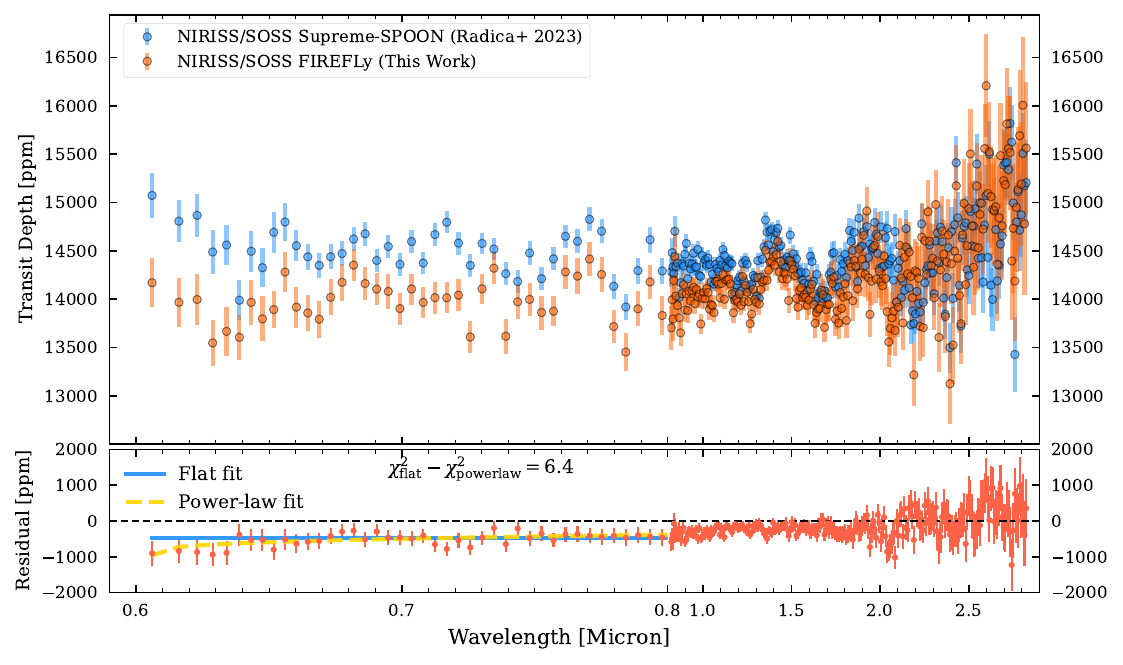}
    \caption{\textit{Top}: SOSS spectrum extracted with FIREFLy compared with that published by \cite{2023MNRAS.524..835R} using Supreme-SPOON. \textit{Bottom}: the residual between the SOSS spectrum extracted with FIREFLy and that published by \cite{2023MNRAS.524..835R} using Supreme-SPOON. The slope at the blue end of the spectrum extracted by \cite{2023MNRAS.524..835R} is not evident in the spectrum extracted with FIREFLy. This is also shown by the slope of the residual at the blue end in the Bottom panel. We fit a power-law (blue in the bottom panel) to the residuals and compare it with a flat-line fit (yellow in the bottom panel) at wavelengths $< 0.8~\mu$m; the power-law residual shape demonstrates a better fit, with a $\Delta\chi^2$ of 6.4.}
    \label{fig:vs radica}
\end{figure*}

A comparison between the spectrum extracted by FIREFLy and that published by \cite{2023MNRAS.524..835R} with Supreme-SPOON\footnote{Now \href{https://github.com/radicamc/exoTEDRF/tree/main}{exoTEDRF}} is shown in Figure \ref{fig:vs radica}. The most prominent difference is that the spectrum from FIREFLy does not have a slope at the blue end. \cite{2023MNRAS.524..835R} attributes the slope to either enhanced Rayleigh scattering or pressure-broadened wings from sodium. However, such a slope is not seen in the VLT data from \cite{2018Natur.557..526N} nor the ACESS spectrum from \cite{2022AJ....164..134M}, so it is not likely that the slope is from the sodium wing. We investigate the cause of such difference in Section \ref{sec:comparison}.


\subsection{Limb Asymmetry}\label{sec:asymmetry}
Planetary atmospheres are not perfectly uniform. The circulation of heat from the irradiated dayside to the non-irradiated nightside can lead to differences in atmospheric properties -- such as temperature, chemical abundances, and cloud coverage -- between the morning and evening terminators of a transiting exoplanet. This phenomenon is referred to as limb asymmetry \citep{caldas2019,ehrenreich2020, espinoza2024}. In hot, tidally locked planets, these asymmetries manifest because the leading limb (often the morning limb) and the trailing limb (evening limb) experience different thermal and dynamical conditions (e.g. wind, atmospheric circulations), leading to slight variations in their effective radii. This discrepancy distorts the transit light curve, causing an apparent shift in the planet’s time of conjunction \citep{dobbs-dixon2012} and creating a degeneracy between true orbital timing and atmospheric asymmetries.

Early detections of limb asymmetries use high-resolution ground-based spectroscopy that isolates the morning and evening terminators by their Doppler-shifted spectral features (e.g. \citealt{ehrenreich2020, bourrier2020,kesseli2021}). JWST enables us to probe limb asymmetries directly by revealing the small timing variations at conjunctions (e.g. \citealt{2023Natur.614..659R}). The morning and evening limb transmission spectra can be extracted from the transit depths extracted at ingress and egress (e.g. \citealt{espinoza2024, murphy2024}), informing us of the differences in cloudiness and molecular abundance between the two limbs.

\begin{figure*}[t]
    \centering
    \plotone{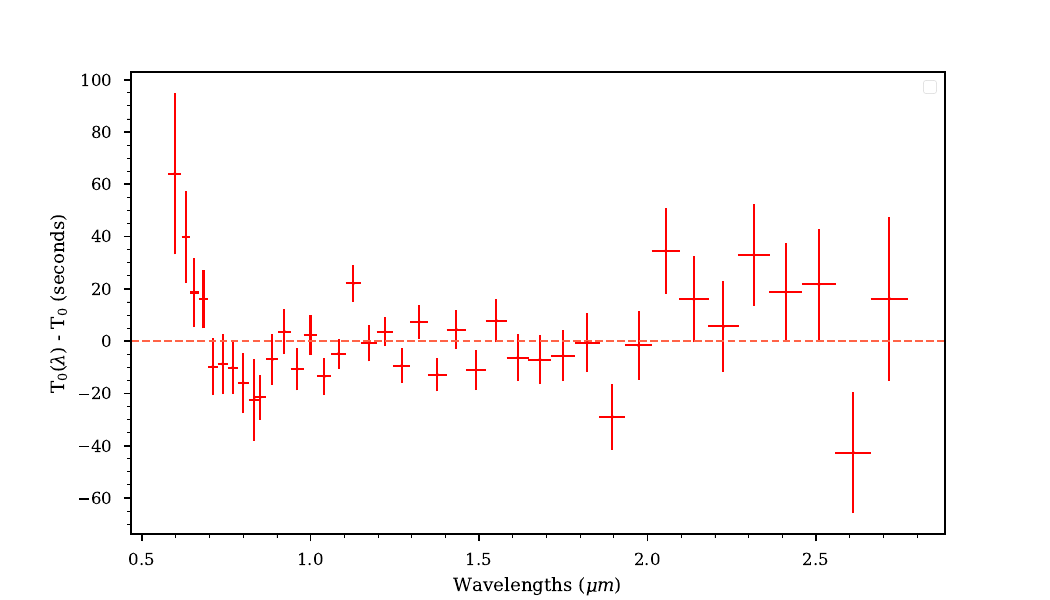}
    \caption{Wavelength dependence of the variation of mid-transit-time $T_0 (\lambda)$ compared to the value fitted for the white light curve at R$=25$. The slope of $T_0$ at the blue end coincides with the sodium wing in the transmission spectrum, suggesting possible day-side and night-side asymmetries in sodium abundance. Other parts of the $T_0$ spectrum are flat with wavelength and do not correlate with spectral features, implying that our extracted abundances are largely unbiased.}
    \label{fig:T0}
\end{figure*}

We investigate the presence of limb asymmetry in WASP-96 b by examining the presence of wavelength dependence on the mid-transit time. We fit $t_0$, $(R_p/R_*)^2$, and the systematics vectors at lower resolution (R$=25$) compared to Figure \ref{fig:transmission spectrum} to reveal the possible $t_0$ variations at higher signal-to-noise. We also extend the bluest wavelength of SOSS order 2 to 0.59 $\mu m$ to get slightly more complete coverage of the sodium wing. The difference between fitted $t_0$ and the mid-transit-time obtained for the white light curve is shown in Figure \ref{fig:T0}. We find a blueward slope that extends to $\sim 0.7$ $\mu m$ on the order of 50 seconds, showing agreement with the pressure-broadened sodium wing revealed by both the VLT data \citep{2018Natur.557..526N} and NIRISS/SOSS data. This slope does not show correlations with our systematic vectors, and visual inspections of the light curves reveal good fits with no significant outliers. This could imply sodium winds being dredged up in the morning side and condensed out in the night side \citep{seidel2023}. As a result, the higher sodium abundance in the evening terminator leads to a positive shift of the mid-transit time. We note that this finding is only tentative as we only have the spectral information for the red side of the sodium wing, and it is degenerate with cloud scattering without the addition of optical short-wavelength data. Follow-up high-resolution RV studies are needed to constrain the possible sodium limb asymmetry fully \citep{louden&wheatley2015,nikolov2015}. Other parts of the $t_0$ spectrum are generally flat with wavelength and do not correlate with spectral features. This implies that the abundance of other species that we extract from our transmission spectrum are largely unbiased to within our errors by the limb asymmetries. In the remainder of this paper, we analyze the transmission spectra obtained with fixed $T_0$ shown in Figure \ref{fig:transmission spectrum}, because the spectral features appear to be symmetric in both limbs, except for the sodium feature that is not fully constrained in the SOSS wavelength range.

\subsection{Atmospheric Retrieval}\label{sec:setup}

\begin{deluxetable*}{lCCCCCCCCC}[t]
\label{tab:priors}
\tablecaption{Priors of Parameters for Our suites of Hierarchical Free and Equilibrium Chemistry Retrievals along with the Bayesian Information Criterion (BIC) obtained for each model. A more negative BIC implies a more preferred model.}
\tablewidth{0pt}
\tablehead{
\colhead{Parameters}& \colhead{Free Prior} &  \colhead{Fixed Value}& \colhead{Model 1}  &  \colhead{Model 2} & \colhead{Model 3} & \colhead{Model 4} & \colhead{Model 5}  & \colhead{Model 6} & \colhead{Equilibrium}
}
\startdata 
$R_{\text{p}}~(R_{\text{jup}})$ & $\mathcal{U}(0.8,1.6)$ &  - &  \text{free} &  \text{free}&  \text{free} &  \text{free} & \text{free} & \text{free} & \text{free}\\
$T_{eq}~(\text{K})$ & $\mathcal{U}(100,3600)$ & - &\text{free}&\text{free}&\text{free}&\text{free}&\text{free}&\text{free} & \text{free}\\
$\log (P_{\text{cloud}})$ (bars)& $\mathcal{U}(-6,2)$ &-&- & \text{free} & \text{free}& \text{free}& \text{free} & \text{free} &\text{free} \\
$\log{f}$ &  $\mathcal{U}(-2,4)$ & 0 & \text{fixed} & \text{fixed} & \text{free} &\text{free}&\text{free}&\text{free} &\text{fixed}\\
$\gamma$ & $\mathcal{U}(-8,1)$ & $-4$ & \text{fixed} & \text{fixed} & \text{free} & \text{free} & \text{free} & \text{free} & \text{fixed}\\
$\theta$ & $\mathcal{U}(0,1)$ & 1 & \text{fixed} & \text{fixed} & \text{fixed} & \text{free} &\text{fixed} &\text{free} & \text{fixed}\\
$T_{\text{int}}~(\text{K})$ & - & 400 &- &- & - &-&\text{fixed} & \text{fixed} & -\\
$\log{\gamma_{\text{O/IR}}}$ & $-\frac{x^2}{8}$, $x\in[0,1]$ & -& -& -& -& -& \text{free} & \text{free} & -\\
$\log{\kappa_{IR}}$ & $\mathcal{U}(-4,2)$ & - & - & - & - &- &\text{free} &\text{free} & - \\
$\text{[Fe/H]}$ & $\mathcal{U}(-1.5,2.5)$ & - & - & - & - & - & - & - & \text{free}\\
$\text{C/O}$ & $\mathcal{U}(0.1,1.6)$ & - & - & - & - & - & - & - & \text{free}\\
$\log{P_{\text{quench}}}$ & $\mathcal{U}(-6,3)$ & - & - & - & - & - & - & - & \text{free}\\
$\log X_{\text{H$_2$O}}$ & $\mathcal{U}(-15,0)$ & - & \multicolumn{6}{C}{\text{free}} & -\\
$\log X_{\text{CH$_4$}}$ & $\mathcal{U}(-15,0)$& - & \multicolumn{6}{C}{\text{free}} & -\\
$\log X_{\text{CO}}$ & $\mathcal{U}(-15,0)$& - & \multicolumn{6}{C}{\text{free}} & -\\
$\log X_{\text{CO$_2$}}$ & $\mathcal{U}(-15,0)$& - & \multicolumn{6}{C}{\text{free}} & -\\
$\log X_{\text{H$_2$S}}$ & $\mathcal{U}(-15,0)$& - & \multicolumn{6}{C}{\text{free}} & -\\
$\log X_{\text{K}}$ & $\mathcal{U}(-15,0)$& - & \multicolumn{6}{C}{\text{free}} & -\\
$\log X_{\text{Na}}$ & $\mathcal{U}(-15,0)$& - & \multicolumn{6}{C}{\text{free}} & -\\
\hline
BIC with SOSS  data & - & - & 470.4 & \textbf{455.1}  & 464.3 & 471.7 & 481.4 & 485.9 & 431.3\\
BIC with combined data & -& -  & 652.3  & \textbf{584.4}  & 597.3 & 613.0  & 618.6 & 643.6 & 571.2 
\enddata
\tablecomments{An additional parameter, the offset between the SOSS and HST/VLT data, is fitted for the joint retrieval with Gaussian prior $\mathcal{N}(0, 200)$ ppm.}
\end{deluxetable*}


We perform an atmospheric retrieval for WASP-96 b with petitRADTRANS (pRT, \cite{Molli_re_2019}). We perform two suites of independent retrievals: \Romannum{1}. a retrieval with SOSS-only data and \Romannum{2}. a retrieval combining SOSS, VLT, HST, and Spitzer. For each suite, we model the atmosphere hierarchically, starting from simple models to complex ones. In particular, six models are explored:
\begin{enumerate}
    \item isothermal and cloud-free model;
    \item isothermal model with a gray cloud deck;
    \item isothermal model with the inclusion of free scattering parameters caused by clouds and hazes;
    \item isothermal model with inhomogeneous clouds and hazes;
    \item  model following the Guillot temperature profile \citep{2010A&A...520A..27G} with the inclusion of cloud and hazes. We fix the internal temperature to 400 K following \citet{Thorngren2019b}.
    \item same model as 5 but with inhomogeneous clouds
\end{enumerate}
The species are retrieved freely while the abundances are assumed to be constant vertically. We include the following species as sources of opacity as they are significant sources of opacity in this spectral region (Table \ref{tab:priors}): H$_2$O, CH$_4$, CO$_2$, CO, H$_2$S, K, and Na. We retrieve the mass fraction of these species with log-uniform prior from $10^{-15}$ to 1. We consider H$_2$-H$_2$ and H$_2$-He collision-induced absorption as sources of gas continuum opacities in addition to H$_2$ Rayleigh scattering. Common to all our retrievals, we fit the following parameters in addition to the mass fraction of the atmospheric species: planet radius ($R_p$), atmospheric temperature ($T$), and an offset of the SOSS spectrum with the combined VLT, HST, and Spitzer spectrum from \cite{2022MNRAS.515.3037N}.

The pRT package uses the correlated-k approach to compute opacities at different wavelengths. The line-by-line data used to create the k-tables are derived from the following sources: H$_2$O \citep{2018MNRAS.480.2597P}, CH$_4$ \citep{2017A&A...605A..95Y,2020ApJS..247...55H}, CO \citep{2010JQSRT.111.2139R,2015ApJS..216...15L}, CO$_2$ \citep{2020MNRAS.496.5282Y}, H$_2$S \citep{2016MNRAS.460.4063A}, K \citep{2019A&A...627A..67M}, and Na \citep{2019A&A...628A.120A}. We use pRT to create forward models for the planet's atmosphere, and the nested sampling algorithm (\citealt{10.1063/1.1835238}) through \verb|PyMultiNest| (\citealt{2009MNRAS.398.1601F,2014A&A...564A.125B}) to estimate the model evidence and the parameter posterior distributions in the Bayesian framework. The retrievals are done with 2000 live points.

We model the cloud and haze following the same prescription as \cite{2008A&A...481L..83L}. We include a power-law slope and enhanced Rayleigh scattering as a proxy for high-altitude hazes, and an opaque cloud deck at pressures $P\geq P_{\text{cloud}}$. The opacity is parameterized by

\begin{align}
    \kappa = \begin{cases}
    f\cdot\kappa_0(\frac{\lambda}{\lambda_0})^\gamma, &(P < P_{\text{cloud}})\\
    \infty, &(P\geq P_{\text{cloud}})
    \end{cases} \label{eq:cloud}
\end{align}

\noindent where $\lambda_0$ is the reference wavelength ($0.35~\mu m$ here), $\kappa_0$ is H$_2$-Rayleigh scattering cross section at $\lambda_0$ ($1.59 \times 10^{-3}$ cm$^2$/g), $\gamma$ is the power-law scattering slope, and f is the Rayleigh scattering enhancement factor. We follow the prescription from \cite{2016ApJ...820...78L} to model inhomogeneous/patchy clouds: the atmosphere is computed with a linear combination of a cloudy atmosphere and a clear atmosphere, with the former being weighted by the cloud fraction $\theta$ and the latter being weighted by $1-\theta$. 
In sum, a list of priors used for all six models is shown in Table \ref{tab:priors}.

We perform an additional retrieval that assumes chemical equilibrium to obtain the carbon-to-oxygen ratio (C/O) and atmospheric metallicity of WASP-96 b. The setup follows the model 2 preferred by the free retrieval (the rightmost column of Table \ref{tab:priors}). We allow $R_p$, equilibrium temperature, metallicity ([Fe/H]), C/O, cloud-top pressure $P_{\text{cloud}}$,  and $P_{\text{quench}}$ (i.e. the pressure below which the abundances of atmospheric species are taken to be constant) to vary, and included the same set of atmospheric species as the free retrieval.


\section{Retrieval Results}\label{sec:result}
We select the best atmospheric models among the six listed in Section \ref{sec:setup} for the two suites of retrievals by comparing the Bayesian evidence for each model. This comparison is interpreted on the frequentist “sigma” scale following \cite{throngren2025}. Specifically, assuming Gaussian errors, the BIC is computed as
\begin{equation}
BIC = \chi^2 + k \ln N
\end{equation}
where $\chi^2$ is the best-fit chi-squared, $k$ is the number of free parameters, and $N$ is the number of data points. The difference $\Delta BIC$ between models quantifies the significance of the removed species. Specifically,
\begin{equation}
    \Delta BIC = BIC_i - BIC_{\text{full}},
\end{equation}
where $BIC_{\text{full}}$ is the BIC for the full free chemistry model and $BIC_i$ is the BIC for the model without the species $i$. Therefore, a larger and more positive $\Delta BIC$ implies stronger evidence for the presence of the species. 

$\Delta BIC$ is translated to a detection significance $\sigma$ via
\begin{equation}
p = \exp\left(-\frac{1}{2}\Delta BIC\right), \quad \sigma = \Phi^{-1}\left(1 - \frac{p}{2}\right)
\end{equation}
where $\Phi^{-1}$ is the standard normal quantile function. 

We find that for both retrieval suites, the isothermal model with only a global low-altitude gray cloud deck is favored
compared to more complex models by more than $\sim$ 3$\sigma$ (Table \ref{tab:priors}). Moreover, the fitted parameters for these more complex models tend to match the values expected for an isothermal, haze-free atmosphere and are well constrained, indicating that the models naturally converge to the simpler physical assumptions. We therefore adopt this best model to infer the detection significance of different species in the WASP-96 b's atmosphere. The inference process involves running the retrieval with the best model without a specific species, and then comparing the Bayesian Information Criterion (BIC) of this model to that of a reference model that includes all species.

In the following sections, we present the retrieved atmospheric parameters for the two suites of independent retrievals based on SOSS-only data and combined SOSS, VLT, HST, and Spitzer data respectively.

\subsection{SOSS-only Retrieval}

\begin{figure*}[t]
    \centering
    \plotone{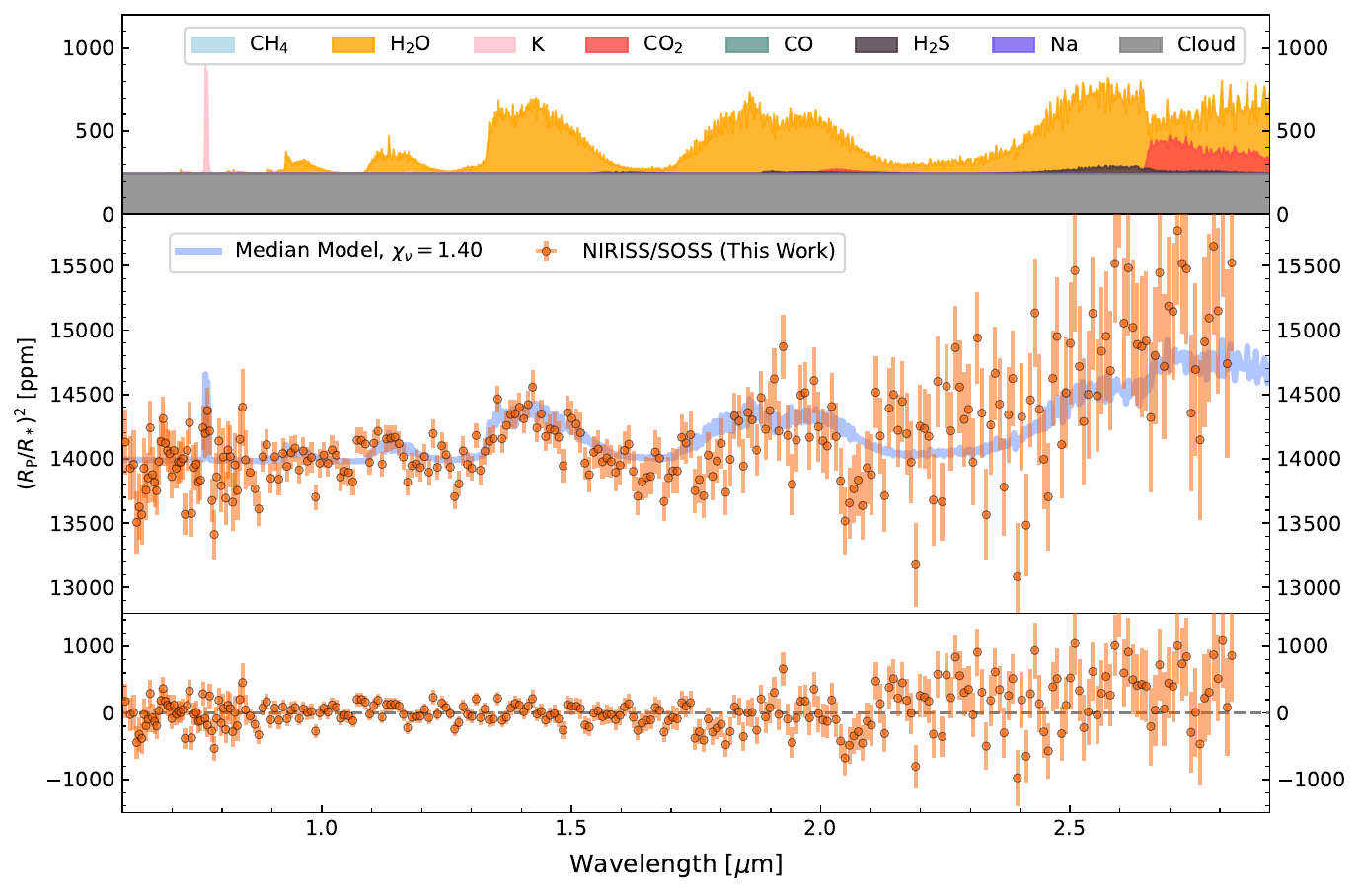}
    \caption{\textit{Top}: The contribution of different chemical species to the observed transmission spectrum. \textit{Middle}: median best-fitting model spectrum with petitRADTRANS, retrieved with SOSS-only data. \textit{Bottom}: the residuals of the spectrum fit. The SOSS-only data constrains the water and potassium features, while the sodium feature is not constrained due to the incomplete wavelength coverage. The retrieval favors the presence of a homogeneous gray cloud deck at low altitudes. The presence of haze is not favored.}
    \label{fig:retrieval_spec_soss}
\end{figure*}

For the free retrieval using only SOSS data, we present the median transmission spectrum at 10-pixel resolution for the isothermal model with a gray cloud deck from pRT - the most favored model among the six - in the middle panel of Figure \ref{fig:retrieval_spec_soss}, and the contributions of different species that explain this spectrum as well as the residuals are shown in the top and bottom panels of the figure respectively. The median model fits the transmission spectrum with $\chi_{\nu} = 1.40$ (DoF = 267). We summarize the retrieved abundance for individual species we tested in Table \ref{tab:species_combined} along with the detection significance on the frequentist 'sigma' scale. The contribution of each species to the observed transmission spectrum is shown on the top panel of Figure \ref{fig:retrieval_spec_soss}.

In summary, water is confidently detected across the SOSS wavelength range. CO$_2$ and CO are disfavored by a small margin of BIC, and interestingly, the SOSS data points do not constrain sodium abundance, which is strongly present in the VLT data. However, one can clearly see an upward slope at the bluest wavelength range in the residual of the spectrum fit (bottom panel of Figure \ref{fig:retrieval_spec_soss}), which the model does not capture because of the limited number of data points. Potassium is marginally preferred by the SOSS data, and its abundance is well-constrained.  As an additional test, we do not find evidence of lithium at $\sim$0.67 $\mu m$ in the extracted transmission spectrum. 

As described above, our hierarchical atmospheric retrieval favors an isothermal atmospheric model with a homogeneous gray cloud deck at an equilibrium temperature of $1248^{+131}_{-132}$ K. The isothermal PT profile is favored over the Guillot profile for WASP-96 b at $> 5\sigma$. A gray cloud deck is favored over a completely cloudless model at 3.7$\sigma$. The model that includes inhomogeneous clouds suggests the majority of the planetary terminator is covered by the cloud ($\theta = 0.96^{+0.03}_{-0.05}$). We find the cloud deck is constrained at relatively low altitudes, with $\log{\text{$P$}_{\text{cloud}}} = -1.56^{+0.41}_{-0.97}$. We do not find evidence of hazes: for the model that includes hazes, we find the scattering slope is largely Rayleigh-like ($\gamma = -4.51^{+2.72}_{-2.31}$), with no evidence of enhancement ($\log{f} = -0.72^{+1.00}_{-0.81}$). The full corner-plot of this model is shown in Figure \ref{fig:soss_only_retrieval}.

For the chemical equilibrium retrieval model using only SOSS data, we use the isothermal model with a gray cloud deck informed by the free retrieval. We find C/O $= 0.60^{+0.12}_{-0.20}$ and [Fe/H] $= 0.07^{+0.67}_{-1.03}$. In comparison, the solar value of C/O is roughly 0.55 \citep{lodders2002}. Our result indicates that the C/O ratio and metallicity of WASP-96 b informed by SOSS data is consistent with solar values within 1$\sigma$.
We note that chemical equilibrium should be a reasonable approximation in this case, as the spectra is dominated by H$_2$O, which is insensitive to vertical mixing and photochemical effects at these temperatures \cite{mukherjee2024}. Moreover, the lower BIC obtained for the equilibrium retrievals compared to the best free retrievals (Table \ref{tab:priors}) also indicates that chemical equilibrium is a reasonable approximation.

\begin{deluxetable*}{l|CCCCCC|CCCCCC}[t]
\label{tab:species_combined}
\tablecaption{Atmospheric inference from NIRISS/SOSS-only and joint VLT, HST, NIRISS/SOSS, and Spitzer observations. We show posterior $\log X_i$ and $\log(\mathrm{VMR})$, along with Bayesian Information Criterion (BIC) comparison for models with and without that species as well as the $\chi^2$ and Degree of Freedom (DoF) for the retrievals without that species. $\Delta BIC =  BIC_{i} - BIC_{\text{full}} $, where $BIC_{\text{full}}$ is the BIC for the full free chemistry model and $BIC_{i}$ is the BIC for the model without the species $i$.}
\tablewidth{0pt}
\tablehead{
\multicolumn{1}{c|}{Species} &
\multicolumn{6}{c|}{\textbf{SOSS}} &
\multicolumn{6}{c}{\textbf{Joint}} \\
\cline{2-7} \cline{8-13}
& \colhead{$\log X_i$} & \colhead{$\log(\mathrm{VMR})$} & \colhead{$\Delta BIC$} & \colhead{$\chi^2$} & \colhead{DoF} & \multicolumn{1}{c|}{Significance} 
& \colhead{$\log X_i$} & \colhead{$\log(\mathrm{VMR})$} & \colhead{$\Delta BIC$} & \colhead{$\chi^2$} & \colhead{DoF} & \colhead{Significance}
}
\startdata
H$_2$O      & $-3.08^{+0.64}_{-0.45}$ & $-3.98^{+0.64}_{-0.45}$ & $+127.5$ & $546.8$ & 287& $+11.1\sigma$ & $-2.62^{+0.43}_{-0.42}$ & $-3.52^{+0.43}_{-0.41}$ & $+177.8$ & $718.7$ & $405$ &$+13.2\sigma$ \\
K           & $-6.34^{+1.53}_{-1.20}$ & $-7.57^{+1.55}_{-1.20}$ & $+3.6$  & $422.8$ & 287 &$+1.4\sigma$  & $-5.76^{+1.05}_{-1.13}$ & $-7.00^{+1.05}_{-1.14}$ & $+24.7$  &$565.6$ & $405$ & $+4.6\sigma$ \\
CO$_2$      & $-4.11^{+0.89}_{-5.36}$ & $-5.39^{+0.89}_{-5.30}$ & $-5.9$  & $413.4$ & 287 & $-1.9\sigma$  & $-8.50^{+4.30}_{-4.34}$ & $-9.76^{+4.39}_{-4.37}$ & $-8.8$  & $532.2$ & 405& $-2.5\sigma$ \\
CO          & $-8.13^{+4.49}_{-4.43}$ & $-9.19^{+4.48}_{-4.44}$ & $-5.5$ & $413.8$ & 287 &  $-1.9\sigma$  &   $-8.67^{+4.41}_{-4.20}$ & $-9.76^{+4.40}_{-4.19}$ & $+3.0$  &$544.0$ & $405$&$+1.2\sigma$ \\
CH$_4$      & $-6.77^{+2.46}_{-5.37}$ & $-7.65^{+2.49}_{-5.35}$ & $-10.1$  & $409.2$ & 287 & $-2.7\sigma$  & $-4.44^{+0.60}_{-3.96}$ & $-5.29^{+0.60}_{-4.00}$ & $+16.9$  & $557.9$& $405$& $+3.7\sigma$ \\
H$_2$S      & $-3.90^{+1.15}_{-7.02}$ & $-5.10^{+1.17}_{-7.02}$ & $-6.7$  & $412.6$& 287 & $-2.1\sigma$  & $-3.31^{+0.83}_{-6.98}$ & $-4.48^{+0.82}_{-6.98}$ & $-11.5$  &$529.5$ & 405& $-3.0\sigma$ \\
Na          & $-9.90^{+3.30}_{-3.31}$ & $-10.90^{+3.34}_{-3.33}$ & $-9.3$ & $410.0$ &  287& $-2.6\sigma$  & $-3.40^{+0.90}_{-0.92}$ & $-4.41^{+0.91}_{-0.92}$ & $+79.1$ & $620.1$ & $405$ & $+8.6\sigma$ \\
Cloud deck  & - & - & $+16.8$ & $436.0$ & 287& $+3.7\sigma$ & - & - & $+67.9$ & $608.9$ & 405& $+8.0\sigma$ \\
\enddata
\tablecomments{Negative `sigma' values represent the case where having a species is disfavored.}
\end{deluxetable*}


\subsection{Joint Retrieval}
\begin{figure*}[t]
    \centering
    \plotone{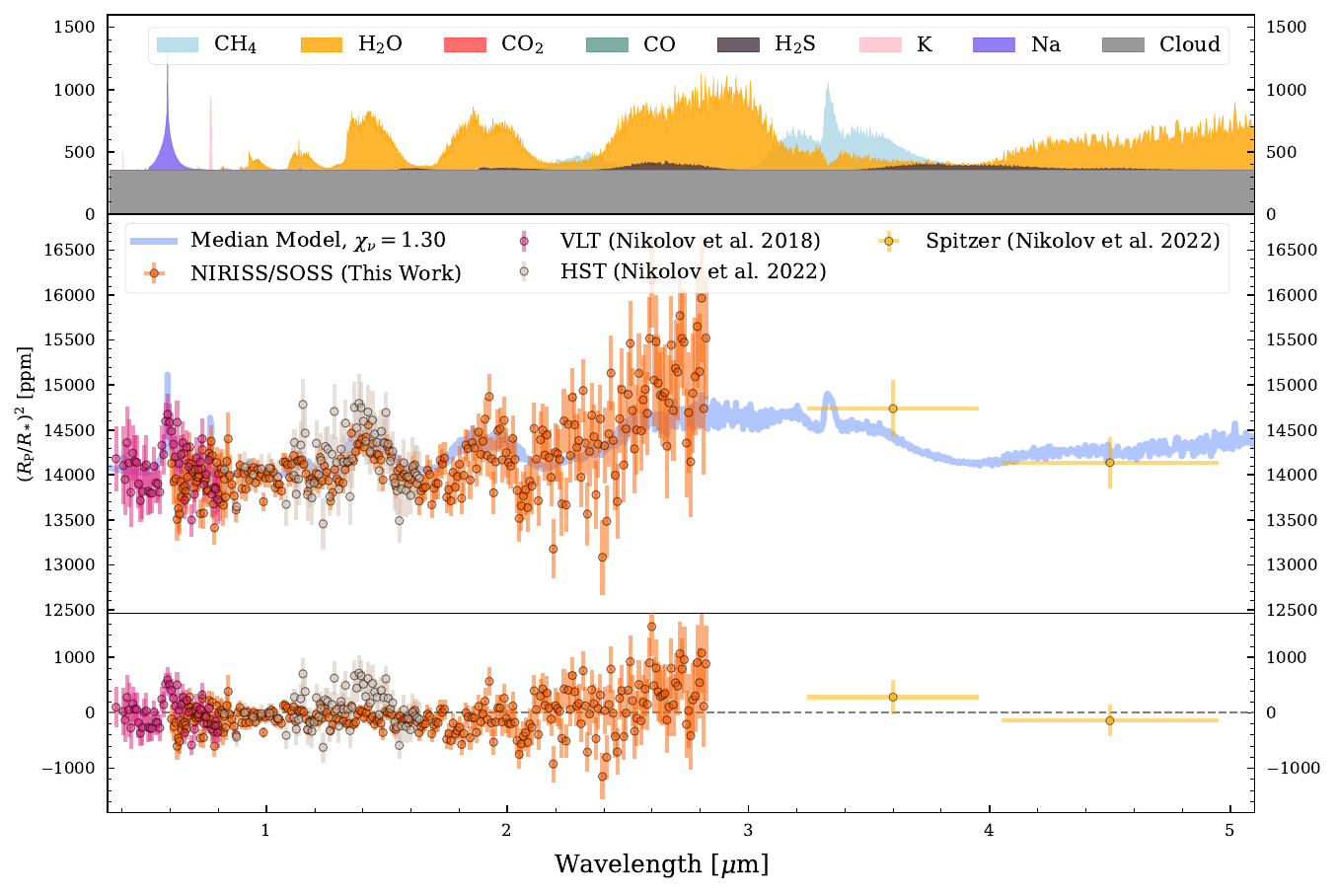}
    \caption{Same a Figure \ref{fig:retrieval_spec_soss}, but for combined SOSS, VLT, HST, and Sptizer spectrum. The combined data constrains sodium, potassium, and water. The presence of methane is not reliable as the model attempts to fit the two photometric points from Spitzer. Same as SOSS-only retrieval, we do not find evidence of haze, and we detect a homogeneous gray cloud deck at low altitudes.}
    \label{fig:retrieval_spec_joint}
\end{figure*}

For retrieval using joint VLT, HST, Spitzer, and SOSS data, we present the median transmission spectrum for the most favored model, the isothermal model with a gray cloud deck, in the middle panel of Figure \ref{fig:retrieval_spec_joint}. The median model fits the combined spectra with $\chi_{\nu} = 1.30$ (DoF = 385). We summarize the retrieved abundance and detection significance for each species for the combined spectra in Table \ref{tab:species_combined}. The contribution of each species to the observed transmission spectrum is shown on the top panel of Figure \ref{fig:retrieval_spec_joint}.

Similar to the retrieval using only SOSS data, water and potassium are detected in the combined spectrum. Because of additional data points at the overlapping wavelengths provided by HST and VLT, the detection confidence of water and potassium is higher. While the model does not favor the detection of sodium with only SOSS data, the combined SOSS and VLT data at around 0.6 $\mu m$ fully constrains the sodium wing at 8.6$\sigma$ confidence. CO is favored by the model, despite the low retrieved median mass ratio, and CO$_2$ is again disfavored. Surprisingly, CH$_4$ is constrained in the combined spectrum, with a detection significance of 3.7$\sigma$. A closer examination of the top and middle panel of Figure \ref{fig:retrieval_spec_joint} shows the model attempts to fit a bump presented by the broad Spitzer photometric point at 3.6 $\mu m$, which corresponds to the CH$_4$ feature favored by the model. The posterior of $\log{(X_{\text{CH}_4})}$ also exhibits a long tail towards the low abundances (Table \ref{tab:species_combined}), further suggesting the disfavoring of CH$_4$. The model also attempts to fit the low Spitzer point at 4.5 $\mu m$, which explains the low median CO$_2$ mass ratio compared to the SOSS-only retrieval as well as the broad uncertainty of mass ratios of CO and CO$_2$.  Indeed, we performed an additional retrieval removing the Spitzer data, and the retrieved abundances for the carbon-bearing species are closer to the retrieval with SOSS-only data ($\log{X}_{\text{CO}} = -8.05^{+4.16}_{-4.05}$, $\log{X}_{\text{CO}_2} = -3.51^{+0.64}_{-1.44}$, and $\log{X}_{\text{CH}_4} = -4.70^{+0.74}_{-5.56}$). We therefore caution that a robust constraint of CH$_4$, CO, and CO$_2$ requires higher resolution data at wavelengths greater than 3 $\mu m$. This highlights the importance of the SOSS synergy with JWST NIRSpec/G395H data, which can provide a panchromatic transmission spectrum from 0.6 $\mu m$ to 5 $\mu m$.

Consistent with the SOSS-only retrieval, our hierarchical atmospheric retrieval for the combined VLT, HST, Spitzer, and SOSS spectrum favors an isothermal atmospheric model with a homogeneous gray cloud deck at equilibrium temperature of $1265^{+125}_{-116}$ K. An isothermal PT profile is favored over the Guillot profile at $>5\sigma$. The joint retrieval tells the same story of the cloud and haze of WASP-96 b as the SOSS-only retrieval: an atmosphere with a global homogeneous ($\theta = 0.97^{+0.02}_{-0.04}$) gray cloud deck ($8.0\sigma$) at low altitudes ($\log{\text{$P$}_{\text{cloud}}} = -1.52^{+0.41}_{-0.86}$) and no evidence of hazes ($\gamma = -4.59^{+2.99}_{-2.32}$ and $\log{f} = -1.00^{+0.80}_{-0.65}$).  The full corner-plot of the model that includes inhomogeneous clouds and hazes is shown in Figure \ref{fig:joint_retrieval}.

For the chemical equilibrium retrieval using combined data, we find solar C/O and metallicity (C/O $= 0.57^{+0.07}_{-0.12}$ and [Fe/H] $= 0.01^{+0.46}_{-0.52}$). This is consistent with retrieval using only SOSS data but with tighter constraints.

\section{Stellar Characterization}\label{sec:stellar}

\begin{figure}[ht!]
    \centering
    \includegraphics[width=1\linewidth]{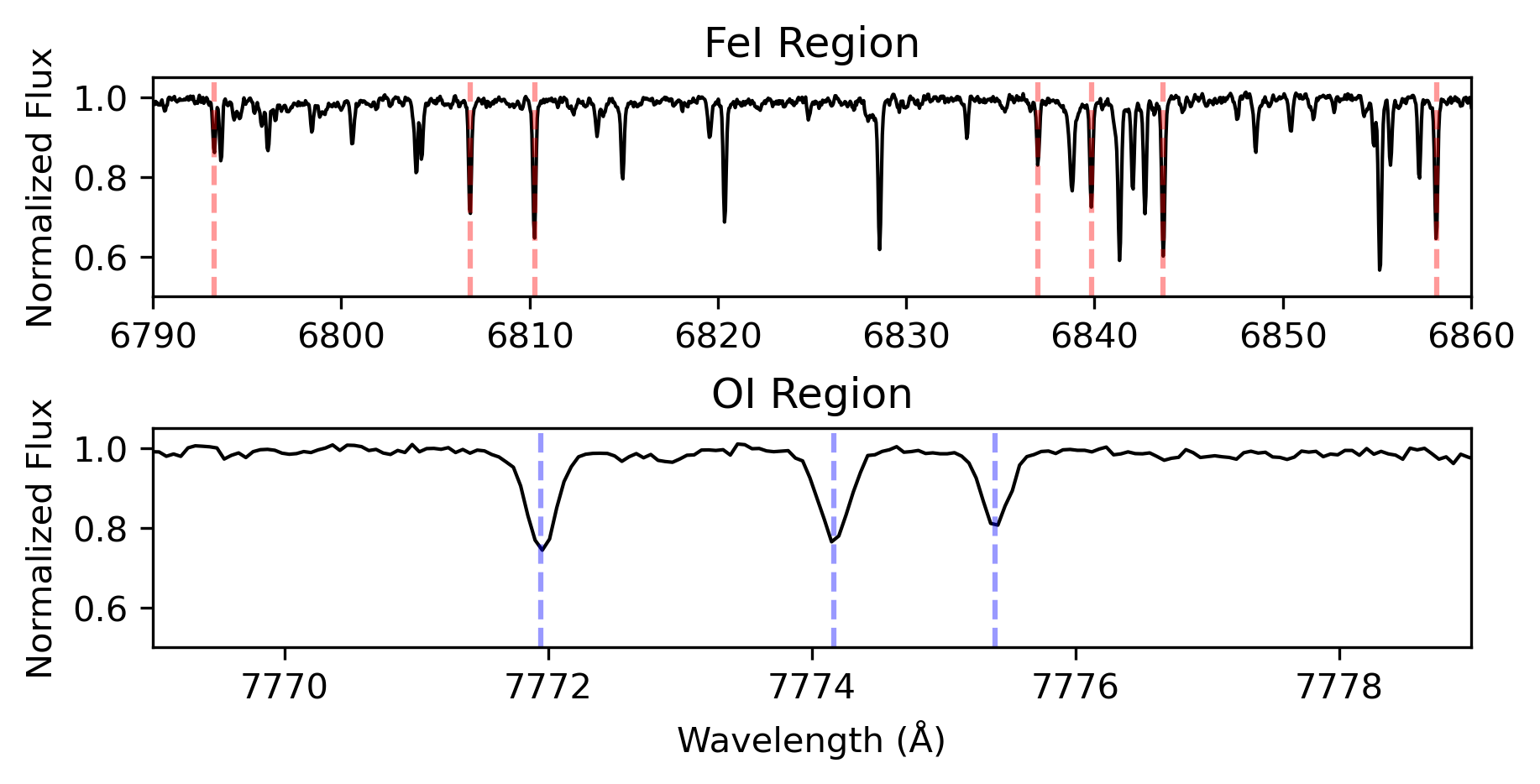}
    \caption{Sample regions of the WASP-96 Magellan/MIKE spectrum used to infer chemical abundances. The red lines indicate locations of the FeI line centers, while the blue lines indicate locations of OI line centers.}
    \label{fig:w96_spec}
\end{figure}

\begin{figure}[ht!]
    \centering
    \includegraphics[width=1\linewidth]{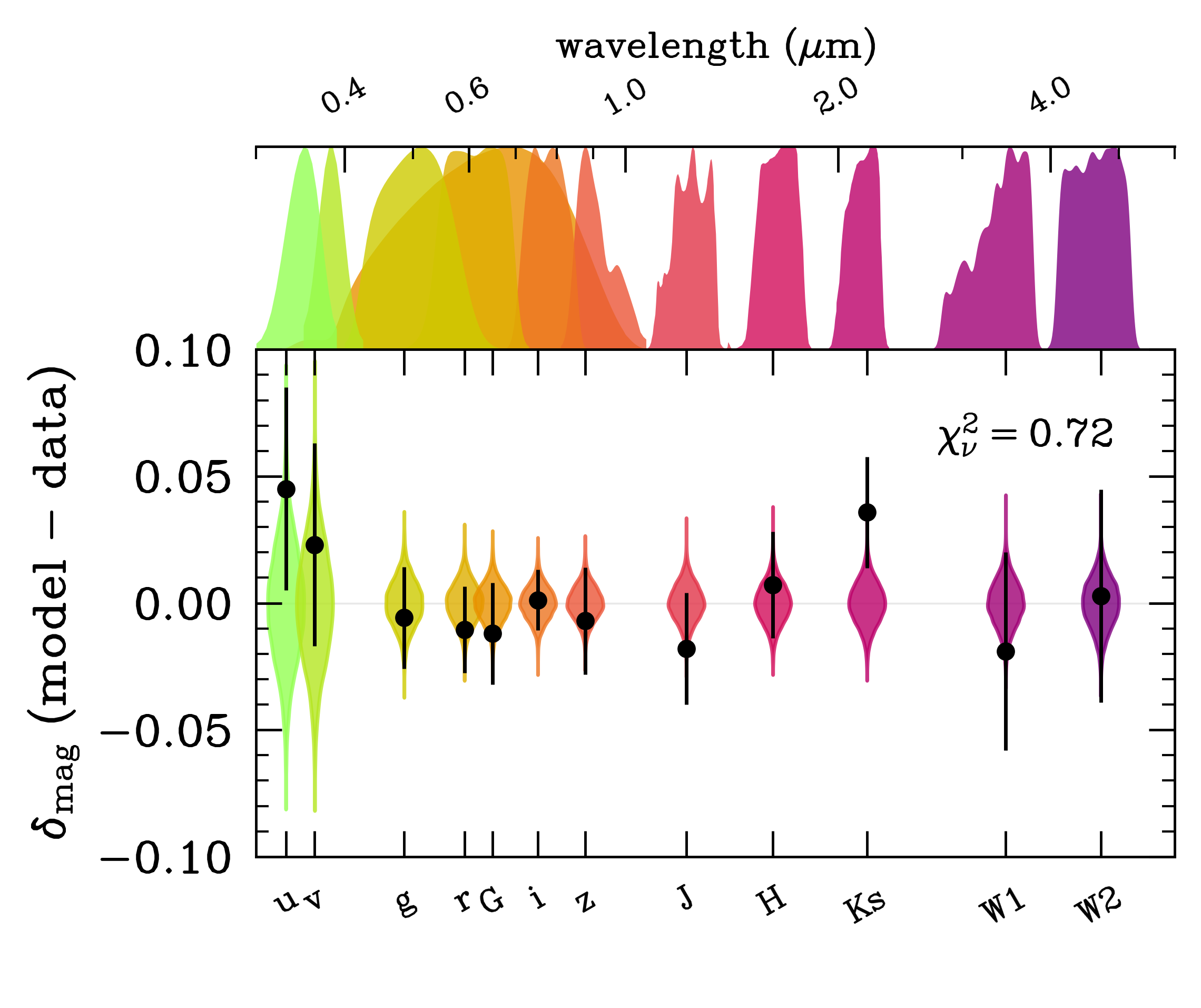}
    \caption{The magnitude fit posteriors of the density-cut \texttt{isochrones} run shows good agreement with the measurements, with $\chi^2_\nu$ = 0.72 indicating our inferences are not biased by photometric offsets.}
    \label{fig:w96_phot_post}
\end{figure}

To characterize the fundamental properties and chemical abundances of the host star WASP-96, we employ a combined methodology based on \citet{Reggiani2022}, \citet{rustamkulov2025}, and \citet{Ross2025}. The approach leverages the precise stellar density derived from JWST transit light curves to refine the stellar age \citep{rustamkulov2025}, while simultaneously constraining stellar parameters and abundances using broadband photometry and high-resolution optical spectroscopy \citep{Reggiani2022}. In brief, we first use stellar model grids to infer preliminary fundamental parameters, including metallicity. A high-resolution optical spectrum is then analyzed to measure equivalent widths from a reference linelist (optical spectrum from Magellan/MIKE; 2018B; PI: M. Alam, see Figure \ref{fig:w96_spec}). A radiative transfer code is used to derive stellar parameters and chemical abundances from these equivalent widths, adopting the photometrically inferred parameters as priors. The resulting spectroscopic posteriors serve as updated priors for a subsequent isochrone fit. This iterative procedure allows for feedback between the photometric-isochrone and spectroscopic analyses until the stellar effective temperature, surface gravity, and metallicity converge consistently between both methods. This yields self-consistent, precise, and accurate stellar parameters and abundances. Finally, the stellar density derived from the JWST light curve, combined with the isochrone posteriors, constrains the stellar mass and therefore age with high precision. The ultra-precise Order 1 white light curve yields a 0.5$\%$-uncertain constraint of WASP-96 b's scaled semimajor axis, $a/R_*$. Assuming a zero-eccentricity orbit for WASP-96 b \citep{Hellier2014}, the best-fit $a/R_*$ yields a measurement of the parent star's bulk density:

\begin{equation}
    \rho_{\star, \, \mathrm{transit}} = \frac{3\pi}{GP^2}\bigg{(}\frac{a}{R_\star}\bigg{)}^3 
    \label{11}
\end{equation}

\noindent where $P$ is the planet's orbital period. Using our values of $a/R_*$ = 8.994$\pm$0.047 and $P = $ 3.4252602 days, we find a stellar mean density of 1.17$\pm$0.018 g cm$^{-3}$, or 0.83 $\pm 0.015 \rho_\odot$. These values are in agreement with other analyses of the same observation \citep{2023MNRAS.524..817T}. Since the eccentricity is measured to be consistent with 0, and the tidal circularization timescale is much shorter than the age of the system, the impact of eccentricity on the stellar density inference is well below the measurement uncertainty. We use this stellar density inference as a constraint on our subsequent fits to the star's fundamental properties, as has been performed in other works \citep[e.g.,][]{Sozzetti2007, Fortney2011, Eastman2023}. Following the procedures described in \citep{rustamkulov2025}, we compile 12 multiband photometric magnitudes spanning NUV to mid-IR wavelengths, the \textit{Gaia} DR3 parallax, a high resolution optical stellar spectrum, and a galactic reddening estimate, which we use as input priors and likelihoods on an \texttt{isochrones} run \citep{morton} (see Table \ref{tab:w96_ins}). The magnitude uncertainties are corrected by their respective zero-point offset uncertainties, and we further increase the $\textit{Gaia G}$, SkyMapper $u$, and SkyMapper $v$ uncertainties to account for the star's possible activity-related 1$\%$-level variability seen in the $Gaia$ $B_\mathrm{p}$ multi-epoch photometry. Using the \texttt{MIST} stellar evolution grid \citep{choi2016}, the \texttt{isochrones} package explores the range of self-consistently allowable stellar parameters that can be explained by the inputs. This approach preserves the complex covariances network shared between the input data and the star's inferred physical parameters. The magnitude-fit posteriors from the \texttt{isochrones} run are shown in Figure~\ref{fig:w96_phot_post} and indicate good agreement with the data ($\chi^2_{\nu} = 0.72$).

We estimate stellar photospheric abundances using a Magellan/MIKE spectrum of WASP-96 obtained on 2018-08-29 04:30:55 UT (2018B, PI: M. Alam, Figure \ref{fig:w96_spec}). Equivalent widths (EWs) of FeI and FeII lines are measured with REvIEW\footnote{https://github.com/madeleine-mckenzie/REvIEW}, using the linelist from \cite{Reggiani2022}. Stellar parameters are then derived with the \textsf{MOOG} wrapper Q2 \cite{Ramirez2014} \footnote{https://github.com/astroChasqui/q2}, which minimizes trends in Fe abundances with reduced EW and excitation potential. Using the posteriors of the inferred photometric solution (see above) in the spectroscopic solutions ensures self-consistent stellar parameters and abundances.

Abundances of non-iron species are determined using REvIEW EWs and the adopted stellar parameters, with NLTE corrections interpolated from precomputed grids. Only elements with multiple reliable lines and valid corrections are reported (Table \ref{tab:w96_abunds}). The uncertainties reported in the table are calculated using the line-by-line scatter. Further details on uncertainty estimation and potential systematics are provided in \cite{Reggiani2024}.

\begin{table}[ht!]
    \centering
    \small
    \caption{WASP-96 Spectroscopically Inferred Photospheric Abundances}
    \begin{tabular}{lccc}
    \hline
    X ($N$) & LTE A(X) & NLTE A(X) & [X/H]$_\text{LTE}$ \\
    \hline
    C I (4) & 8.870 $\pm$ 0.224 & 8.857 & 0.41 \\
    O I (3) & 9.061 $\pm$ 0.058 & 8.892 & 0.37 \\
    Na I (2) & 6.591 $\pm$ 0.019 & 6.480 & 0.37 \\
    Mg I (2) & 7.577 $\pm$ 0.167 &  & 0.03 \\
    Al I (4) & 6.662 $\pm$ 0.034 &  & 0.23 \\
    Si I (14) & 7.776 $\pm$ 0.060 & 7.760 & 0.27 \\
    Ca I (11) & 6.494 $\pm$ 0.054 & 6.269 & 0.19 \\
    Sc II (5) & 4.636 $\pm$ 0.854 &  & 0.21 \\
    Ti I (13) & 5.154 $\pm$ 0.095 &  & 0.18 \\
    Ti II (8) & 5.061 $\pm$ 0.285 &  & 0.09 \\
    V I (8) & 4.210 $\pm$ 0.101 &  & 0.31 \\
    Cr I (13) & 5.986 $\pm$ 0.306 &  & 0.37 \\
    Cr II (4) & 5.753 $\pm$ 0.275 &  & 0.13 \\
    Mn I (2) & 6.728 $\pm$ 0.146 &  & 1.31 \\
    Fe I (46) & 7.684 $\pm$ 0.030 & 7.654 & 0.22 \\
    Fe II (15) & 7.740 $\pm$ 0.069 & 7.738 & 0.28 \\
    Co I (9) & 6.449 $\pm$ 0.401 &  & 1.51 \\
    Ni I (19) & 6.535 $\pm$ 0.057 &  & 0.34 \\
    Cu I (3) & 6.297 $\pm$ 0.719 &  & 2.12 \\
    Y II (2) & 3.197 $\pm$ 0.504 &  & 0.99 \\
    Ba II (2) & 4.123 $\pm$ 0.175 & 4.103 & 1.85 \\
    \hline
    Ratio & Value & Uncertainty & \\
    \hline
    C/O$_\text{NLTE}$ & 0.923 & 0.249 & \\
    $[\text{C/O}]_\text{NLTE}$ & 0.195 & 0.117 & \\
    \hline
    \end{tabular}
    \label{tab:w96_abunds}
\end{table}

\begin{figure}[ht!]
    \centering
    \includegraphics[width=1\linewidth]{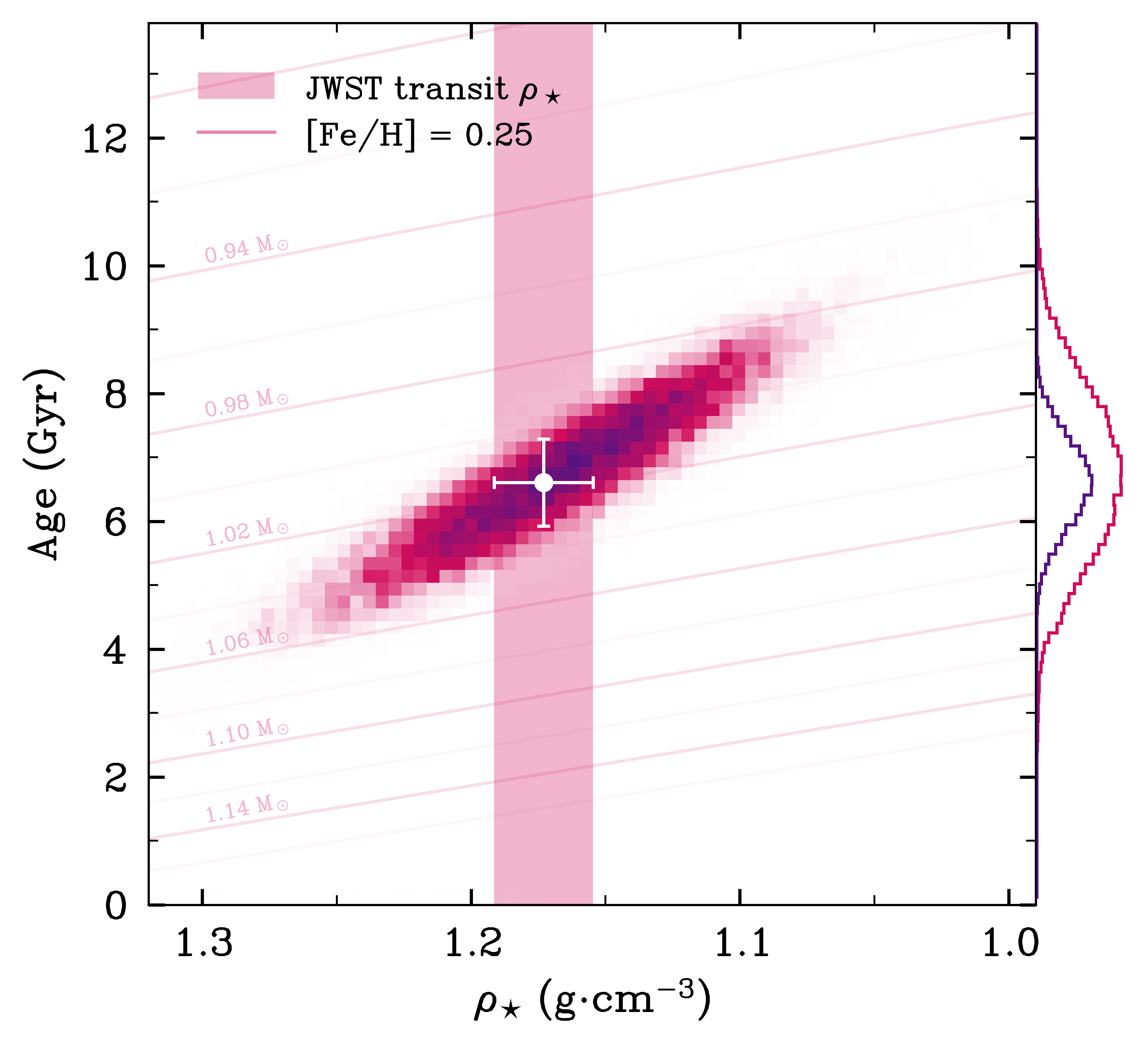}
    \caption{The 2D density-age posterior histogram overlaid with the stellar density (vertical shaded bar) inferred from the transit of WASP-96 b. The MIST evolution tracks for $[$Fe/H$]$ = 0.25 are shown as faint magenta. The marginal posteriors with and without the density cut are shown as magenta and purple histograms, respectively (right).}
    \label{fig:w96_age_post}
\end{figure}

We use our best fit value of $[Fe/H]$ = 0.22 $\pm$0.05 as an input likelihood, and run $\texttt{isochrones}$ with a \texttt{MultiNest} fit using 7000 live points and an evidence tolerance of 0.1. Then, we apply the transit-derived density of 1.17$\pm$0.018 g cm$^{-3}$ as a cut to the posterior using Gaussian elimination (see Figure \ref{fig:w96_age_post}). The $\textit{JWST}$ density constraint is shown in Figure \ref{fig:w96_age_post} as a vertical shaded region.

\begin{table}[ht!]
		\centering
		\begin{tabular}{cccccccccccc}
			\hline 
			Parameter & Unit & Prior \\
			\hline
			$A_V$ & mag & $\mathcal{U}(0.0, 0.1)$ \\
			distance& pc &  $\mathcal{U}(340, 350)$ \\
			age & $\log_{10}$(yr) & $\mathcal{U}(7, 10.1374)$ \\

            \hline

    Parameter & Unit & Likelihood\\
			 \hline
    $A_V$ & mag & 0.03 $\pm$ 0.03 \\
	$[$Fe/H$]$ & & 0.24 $\pm$0.05  \\
    parallax & mas & 2.879531 $\pm$0.02\\

     \hline
        Bandpass & Wavelength range ($\mu$m) & Magnitude \\
			 \hline
            SkyMapper $u$ & 0.31-0.38 & 14.469 $\pm$ 0.04\\
            SkyMapper $v$ & 0.36-0.42 & 14.093 $\pm$ 0.04\\
            SkyMapper $g$ & 0.43-0.64 & 12.687 $\pm$ 0.02\\
            SkyMapper $r$ & 0.50-0.70 & 12.343 $\pm$ 0.017\\
            SkyMapper $i$ & 0.70-0.86 & 12.192 $\pm$ 0.021\\
            SkyMapper $z$ & 0.84-1.03 & 12.161 $\pm$ 0.021\\
            \textit{Gaia} $G$ & 0.39-0.97 & 12.372 $\pm$ 0.02\\
            2MASS $J$ & 1.1-1.4 & 11.267 $\pm$ 0.022\\
            2MASS $H$ & 1.4-1.8 & 10.936 $\pm$ 0.021\\
            2MASS $K_s$ & 2.0-2.3 & 10.914 $\pm$ 0.022\\
            WISE $W1$ & 2.7-3.9 & 10.844 $\pm$ 0.039\\
            WISE $W2$ & 4.0-5.4 & 10.903 $\pm$ 0.042\\
			\hline 
		\end{tabular}
        
        \caption{The input priors and likelihoods ingested in our \texttt{isochrones} run for WASP-96. For each bandpass, the listed wavelength range encloses the central 80$\%$ of the filter's response curve. All magnitude uncertainties are inflated by their respective zero-point calibration uncertainty.}
		\label{tab:w96_ins} 
	\end{table}

\begin{figure}[ht!]
    \centering
    \includegraphics[width=1\linewidth]{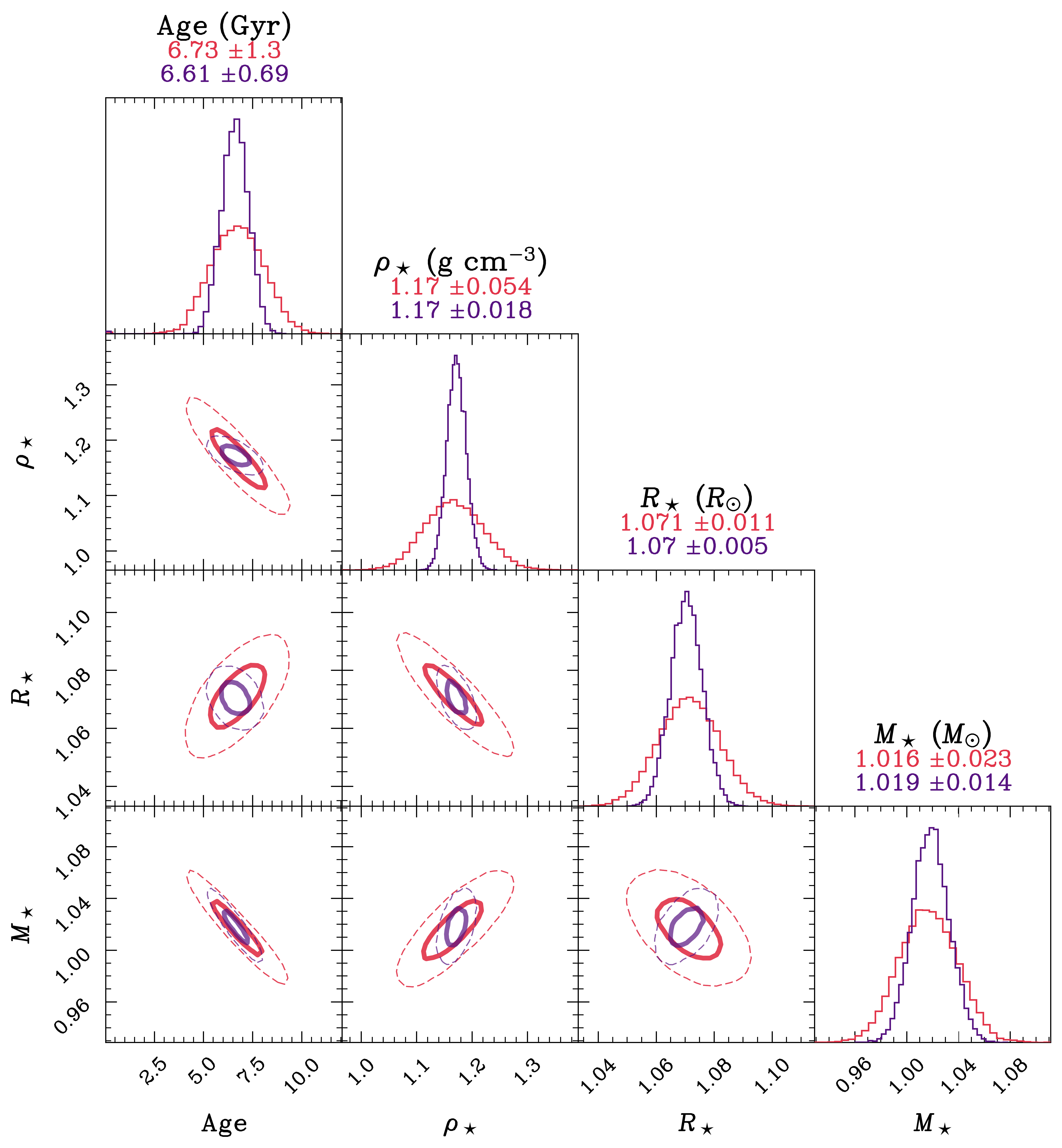}
    \caption{Truncated corner plot showing the \texttt{isochrones} posterior (red) overlaid by the transit density-cut posterior (purple). The transit density cut constrains the strongly covarying mass/radius/age posteriors, reducing their uncertainties by a factor of $\sim$2.}
    \label{fig:w96_corner}
\end{figure}

\begin{table*}
    \centering
    \caption{Updated stellar parameters for WASP-96 with 1-$\sigma$ uncertainties. The density cut did not significantly change the central values of any parameters, and rather reinforced them by collapsing the correlated age, mass, and radius uncertainties.}
    \begin{tabular}{ccccc} \hline
Parameter & Description & Units & With Transit $\mathbf{\rho_\star}$ & $\texttt{isochrones}$ \\ \hline

\textit{Fitted stellar parameters} \\

$\rho_\star$ & density & g cm$^{-3}$ & 1.173 $\pm$ 0.018 & 1.172 $\pm$ 0.05 \\
$M_\star$ & mass & M$_\odot$ & 1.018 $\pm$ 0.014 & 1.016 $\pm$ 0.023 \\
$R_\star$ & radius & R$_\odot$ & 1.070 $\pm$ 0.005 & 1.071 $\pm$ 0.011 \\
log($g$) & surface gravity & $\log_{10}$(cm s$^{-2}$) & 4.387 $\pm$ 0.006 & 4.386 $\pm$ 0.015 \\
$T_\mathrm{eff}$ & effective temperature & K & 5587 $\pm$32 & 5585 $\pm$41 \\
$L_\star$ & luminosity & $L_\odot$ & 1.005 $\pm$ 0.025 & 1.006 $\pm$0.025 \\
$\text{[Fe/H]}$ & metallicity & dex & 0.26 $\pm$ 0.05 & 0.26 $\pm$ 0.05 \\
$d$ & distance & pc & 347.2 $\pm$ 2.2 & 347.3 $\pm$ 2.4 \\
$A_V$ & $V$-band extinction & mag & 0.03 $\pm$ 0.03 & 0.03 $\pm$ 0.03 \\
$\tau_{\mathrm{iso}/\rho}$ & transit age\textsuperscript{$\star$} & $\times$10$^9$ yr & 6.6 $\pm$ 0.7 & 6.7 $\pm$ 1.3 \\
\hline
    \end{tabular}
    \\
    \footnotesize{\textsuperscript{$\star$}Transit density constrained isochronal age.}
    \label{tab:w96_params}
\end{table*}

The best-fit parameters resulting from the density cut are shown in Table \ref{tab:w96_params}, which also shows the same parameters without a density cut applied. The density cut does not shift any of the central values away, while improving the uncertainties in the degenerate parameters $M_\star$, $R_\star$, $T_{\mathrm{eff}}$, and the isochronal age, $\tau_{\mathrm{iso}/\rho}$ by a factor if $\sim 2$ (Figure \ref{fig:w96_corner}). Namely, we find $R_\star = 1.07 \pm 0.005 R_\odot$, $M_\star = 1.018 \pm 0.014 M_\odot$, $T_\mathrm{eff} = 5587 \pm 36$ K, $L_\star = 1.005 \pm 0.025 L_\odot$, and an age of 6.6 $\pm$0.7 Gyr. These values, and our super-Solar $[$Fe/H$]$, are in agreement with the previous findings of \cite{Hellier2014} and \cite{Bonomo2017}. Our reported random uncertainties do not take into account systematic effects such as the star's unknown helium fraction and convective overshoot of the core, nor the systematic differences between stellar models and the uncertainties inherent to their input physics \citep[][i.e.]{Tayar2022}. Marginalizing over these systematic uncertainties, we estimate systematic age and mass uncertainties of 1 Gyr and 2$\%$, respectively. Within these uncertainties, we confirm the status of WASP-96 as a Solar analog, with an 80$\%$ higher metal content. WASP-96 is slated for 3 back-to-back sectors of photometric monitoring by TESS in Summer 2026, which may help refine its rotation period, thus constraining the tidal history shared among the star and planet.

\section{Discussion}\label{sec:discussion}
\subsection{Comparison with Earlier Results}\label{sec:comparison}
Compared to the reduction provided by \cite{2023MNRAS.524..835R}, our reduced spectrum does not have a slope at the blue end that extends a long wavelength range, as shown in Figure \ref{fig:vs radica}. This updated result, however, agrees with the ground-based VLT data from \cite{2018Natur.557..526N}. Consequently, our reduced transmission spectrum has different interpretations compared to that from \cite{2023MNRAS.524..835R} and \cite{2023MNRAS.524..817T}. Firstly, to explain the slope at the blue end, \cite{2023MNRAS.524..817T} invoked an enhanced Rayleigh scattering slope caused by small aerosol particles, and \cite{2023MNRAS.524..835R} argued the red wing of the pressure-broadened sodium absorption line could also lead to such a slope. As presented in Section \ref{sec:result}, with the same haze parametrization, we do not find evidence of enhanced Rayleigh scattering in our retrieval ($\log{f} = -0.71^{+1.00}_{-0.81}$ in our work vs $\log{f} > 1$ reported by \citealt{2023MNRAS.524..817T}). Additionally, while \citet{2023MNRAS.524..817T} constrained sodium abundance using only SOSS data, our analysis demonstrates that it is unreliable to constrain sodium abundance without including data from shorter wavelength ranges. The reported constraint on sodium by \citet{2023MNRAS.524..817T} is partially driven by the broad slope at the blue end of their spectrum, which increases the confidence in sodium detection. Furthermore, the high gray cloud deck pressure reported by \citet{2023MNRAS.524..817T} ($\log{\text{P}_{\text{cloud}}} > 0$) suggests that their model partially attributes the pressure-broadened sodium wing to explain the observed blue-end slope. Our result constrains the cloud-top pressure at $\log{\text{$P$}_{\text{cloud}}} = -1.56^{+0.41}_{-0.97}$. The value reported by \citet{2023MNRAS.524..817T} is 2 orders of magnitude higher, leading to their conclusion that a gray cloud deck is not present in WASP-96 b's atmosphere. Our result suggests a gray cloud deck is present in WASP-96 b's atmosphere, but at a relatively low altitude and high pressure. Finally, \citet{2023MNRAS.524..817T} reported a 6$\sigma$ preference for inhomogeneous clouds and hazes, while our retrieval finds a global homogeneous cloud deck is sufficient to explain the observed transmission spectrum.

The difference between our most recent results and the previously published SOSS analysis likely arises from differences in the \verb|jwst| pipeline version and calibration files. To investigate this, we attempted to reproduce the transmission spectrum published by \citet{2023MNRAS.524..835R} by re-reducing the uncalibrated WASP-96 b SOSS data using the Supreme-SPOON pipeline, using exactly the same calibration and light curve fitting methods. The resulting transmission spectrum shows a much better agreement with our updated results (Figure \ref{fig:reduction_latest}). We fit a power-law to the residuals of Figure \ref{fig:vs radica} and Figure \ref{fig:reduction_latest} respectively and compare it with a flat-line fit. Comparing our FIREFLy results with those published by \cite{2023MNRAS.524..835R} at wavelengths $< 0.8~\mu$m (Figure \ref{fig:vs radica}), the power-law residual shape demonstrates a better fit, with a $\Delta\chi^2$ of 6.4. This $\Delta\chi^2$ value is 0.7 when comparing the refitted spectrum with the supreme-spoon results from our study and the spectrum produced by FIREFLy (Figure \ref{fig:reduction_latest}). The result implies the need to invoke an enhanced scattering slope to explain the difference. We perform the same test comparing the updated supreme-spoon result from our study and that published by \cite{2023MNRAS.524..835R} in Figure \ref{fig:supreme_spoon_comparison}. A power-law slope provides a better fit with $\Delta\chi^2$ of 3.0. The only difference in the data reduction process between the two versions of the spectra is the version of the \verb|jwst| pipeline and calibration files, indicating that the updated calibration methods in the \verb|jwst| pipeline can likely result in transmission spectra with significantly different interpretations. This finding is also hinted by \cite{liuwang2025}, who performed a reanalysis of JWST NIRISS/SOSS and NIRSpec/G395H data for HAT-P-14 b. Because the \verb|jwst| pipeline and calibration files have undergone numerous updates since \cite{2023MNRAS.524..835R}, isolating which specific change drives the spectral differences reported here is beyond the scope of this work.  We encourage future analyses to examine this effect in detail.

Note that an apparent offset in baseline transit depth is present in Figure \ref{fig:reduction_latest}, but this is expected from two effects. First, orbital-parameter choices: our Supreme-SPOON reduction fixes $a/R_\star$ and $b$ to \cite{2023MNRAS.524..835R} for both orders, whereas we fix them to each order’s best-fit white-light solution (Table \ref{tab:wlc}). Because transit depths covary with $a/R_\star$ and $b$, these differing assumptions naturally shift baseline transit depths, an effect that appears larger in lower-S/N channels. Second, 1/f noise treatment: per Radica, M. (in prep.), applying group-level 1/f removal tends to dilute transit depths at wavelengths $\gtrsim 2~\mu$m. At the time of our reduction (March 2024), FIREFLy did not include group-level 1/f removal, while Supreme-SPOON did; this explains FIREFLy’s slightly higher depths beyond 2 $\mu$m in Figure \ref{fig:reduction_latest} and matches the behavior seen in \cite{2023MNRAS.524..835R} (their Figure C5), where pipelines that does not apply group-level 1/f removal (NAMELESS) sit systematically higher. This phenomenon also implicitly appears in other work (e.g. \citealt{coulombe2023,marylou2024}).

\begin{figure}[t]
    \centering
    \plotone{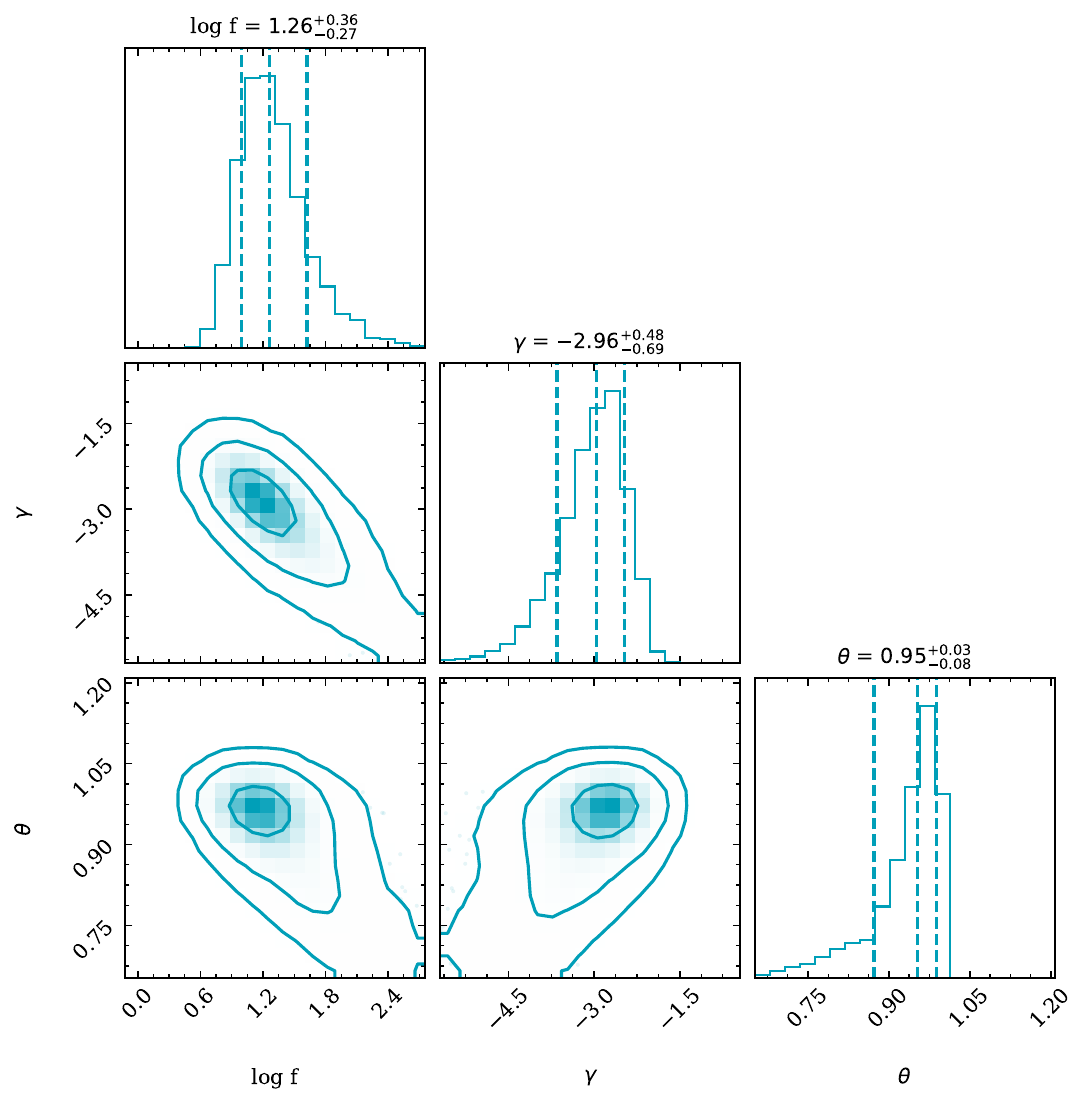}
    \caption{Posterior distributions of the patchiness parameter ($\theta$), haze factor ($\log{f}$), and power-law scattering slope ($\gamma$) using our retrieval framework on the spectrum from \cite{2023MNRAS.524..835R}. The 2D joint posteriors show that patchiness is anti-correlated with the haze factor and positively correlated with $\gamma$, indicating degeneracies between these parameters. These degeneracies may account for the higher retrieved patchiness values compared to \citet{2023MNRAS.524..817T}, as our retrieved haze factor is slightly lower and $\gamma$ is slightly larger.}
    \label{fig:radica_retrieval}
\end{figure}
\begin{figure*}[t]
    \centering
    \plotone{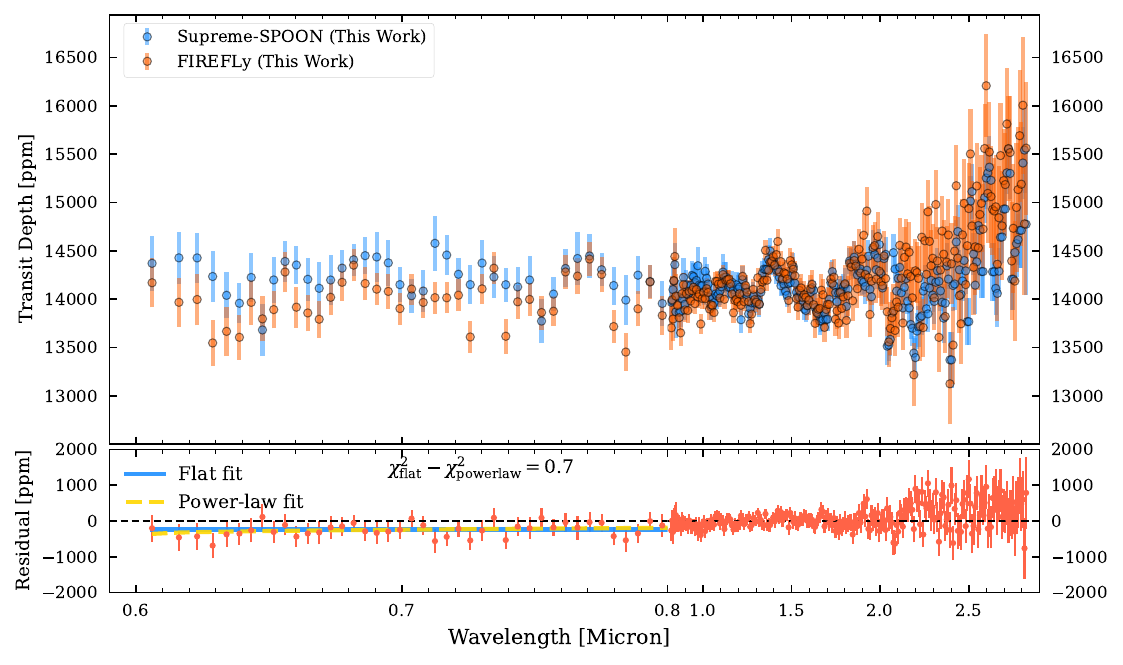}
    \caption{Same as Figure \ref{fig:vs radica}, but comparing the NIRISS/SOSS spectrum reduced by FIREFLy and that reduced by Supreme-SPOON using the most up-to-date JWST calibration methods. \textit{Top}: Compared to Figure \ref{fig:vs radica}, the spectra show better agreement. In particular, the haze-like slope at the blue end reported by \cite{2023MNRAS.524..835R} and \cite{2023MNRAS.524..817T} is not present in the latest reduction. \textit{Bottom}: The residual between the two spectra. The difference between the two reductions at the blue end that is evident in Figure \ref{fig:vs radica} is not prominent when the comparison is made against the latest Supreme-SPOON reduction conducted in this work. We perform the same power-law fit vs flat line fit to the residuals as Figure \ref{fig:vs radica}; while the power-law provides a better fit with $\Delta\chi^2$ of 6.4 in Figure \ref{fig:vs radica}, that statistic is 0.7 in this comparison, indicating the slope feature is not as significant.}
    \label{fig:reduction_latest}
\end{figure*}

We then investigate if the difference in retrieval results could be due to different retrieval codes. we performed an atmospheric retrieval using our petitRADTRANS framework on the original supreme-SPOON spectrum from \cite{2023MNRAS.524..835R}. We use model 4 to retrieve this original spectrum, as it most closely matches the retrieval setup from \cite{2023MNRAS.522.5062B}. The results are broadly consistent with those reported by \cite{2023MNRAS.524..817T}: we recover a significantly enhanced scattering slope parametrized by a haze factor (f) approximately an order of magnitude larger than that retrieved from our updated spectrum and a cloud deck at very low pressure ($\sim 0.29^{+1.06}_{-1.00}$ bar), indicating a negligible contribution from grey clouds. This indicates an enhanced scattering slope has to be invoked to explain the spectrum presented by \cite{2023MNRAS.524..835R}. One notable difference, however, lies in the retrieved patchiness parameter. We find a value of $0.95^{+0.03}_{-0.08}$, which is consistent with their reported result within uncertainties but points toward a more homogeneous atmosphere. Given the minimal role of clouds at such low pressures, this constraint is likely driven by the homogeneity of hazes. Upon closer inspection of the posterior distributions, we find that patchiness is anti-correlated with the haze factor and positively correlated with the power-law scattering slope ($\gamma$) (Figure \ref{fig:radica_retrieval}). Compared to \cite{2023MNRAS.524..817T}, our retrieval yields a slightly lower haze factor and a larger $\gamma$, suggesting that degeneracies between these parameters may contribute to the observed differences in patchiness. It is possible that the subtle differences in model configuration can influence parameters like patchiness, particularly when the atmosphere is close to homogeneous.

Our metallicity retrieved from both the SOSS spectrum ([Fe/H] $= 0.07^{+0.67}_{-1.03}$) and the combined spectrum ([Fe/H] $= 0.01^{+0.46}_{-0.52}$) is consistent with constraints from \cite{2023MNRAS.524..835R} (between 1$\times$ and 5$\times$ solar) and  \cite{2023MNRAS.524..817T} ($\log{Z/Z_{\odot}} =  -0.63^{+0.64}_{-0.44}$), despite slightly larger median values. This result is also consistent with the analysis of pre-JWST data. For example, \cite{2022AJ....164..134M} found a metallicity of $Z/Z_{\odot} = 0.32^{+2.91}_{-0.20}$ and both \cite{2021AJ....161....4Y} and \cite{2022MNRAS.515.3037N} reported a solar to super-solar metallicity.

The possible sodium limb asymmetry that we show in Section \ref{sec:asymmetry} complicates our interpretation of the sodium abundance. Our data is inherently limited to comprehensively examine the limb asymmetry of sodium because 1. the JWST data is not able to entirely capture the sodium wings and 2. the VLT data is limited by precision and telluric-dominated systematics, which could fail to capture the minor time difference at conjunctions caused by the limb asymmetry. Given we only find evidence of limb asymmetry for the sodium feature, our study thus ignores the separate contributions from the morning and the evening limbs. This, however, could bias the retrieved sodium abundances (\citealt{taylor2020}; Mukherjee et al. in prep). Other abundances reported by our study are unbiased by the limb asymmetry. Future studies can use either JWST NIRSpec/PRISM (e.g. \citealt{2023Natur.614..659R}) or high-resolution radial velocity study (e.g. \citealt{ehrenreich2020}) to fully analyze the sodium limb asymmetry hinted by our study.

\subsection{Combining JWST with VLT, HST, and Spitzer}

Our analysis demonstrates a powerful synergy between JWST, VLT, HST, and Spitzer. We find a statistically insignificant offset ($37\pm 24$ ppm) between our SOSS spectrum relative to the combined VLT-HST-Spitzer spectrum published by \cite{2022MNRAS.515.3037N}. Previous studies, including \citet{2021AJ....161....4Y}, \citet{2022MNRAS.515.3037N}, and \citet{2022AJ....164..134M}, have reported offsets between VLT and HST spectra, primarily attributed to the common-mode corrections used to mitigate systematics. In this work, we find that the HST and JWST/NIRISS SOSS spectra exhibit relatively consistent transit depths, suggesting better agreement between these space-based instruments. The fact that the SOSS spectrum simultaneously agrees with the VLT and HST spectrum at the overlapping wavelength regions indicates that while offsets may be needed to align ground-based and space-based instruments due to different systematics, these adjustments do not introduce bias into the interpretation of the transmission spectrum.

Our result shows that the SOSS data alone is not sufficient to constrain sodium. An upward slope at the bluest end in the residual of the spectrum fit due to the sodium wing can be seen (bottom panel of Figure \ref{fig:retrieval_spec_soss}), but the model fails to capture it due to the limited number of data points provided by the SOSS spectrum. However, combining the SOSS data with VLT observations yields a 
8$\sigma$ detection of sodium. This highlights the importance of including additional data at wavelengths bluer than 0.6 $\mu m$ for future studies aiming to use JWST to constrain sodium features effectively.

We caution against the use of Spitzer data to constrain carbon-bearing species, such as CO$_2$, CO, and CH$_4$, due to the limited resolution, which could significantly bias the retrieval results. The detection of CH$_4$ in the combined spectrum is because the model attempts to fit the bump presented by a single Spitzer point at 3.6 $\mu m$. Likewise, the inconsistent CO and CO$_2$ abundances between SOSS-only retrieval and the combined retrieval are primarily due to the model fitting a lower transit depth indicated by the single Spitzer photometric point at  4.5 $\mu m$. A more precise and accurate constraint on these carbon-bearing species requires higher-resolution data between 3 $\mu m$ to 5 $\mu m$, highlighting the importance of the JWST NIRSpec observations for WASP-96 b.

\subsection{Planet Formation Implication}

WASP-96 b uniquely offers an excellent case study for inferring planet formation history based on atmospheric abundances. The constrained refractory species, i.e. sodium and potassium, in a largely clear atmosphere allows us to obtain a refractory-to-oxygen ratio without being hindered by significant clouds. While a grey cloud deck is detected in our retrievals, they are consistently confined to high pressures and therefore do not significantly obscure the photospheric regions probed by the transmission spectrum. The refractory elements remain in solid states in most of the protoplanetary disk, while the volatile content of the accreted planetesimals varies depending on the location of the disk due to its lower condensation temperature.  By measuring refractory-to-oxygen ratio in the planetary atmosphere and compare that ratio to the stellar values, we can obtain an estimate of the rock-to-ice ratio of the accreted planetesimals and determine the atmospheric accretion location of the planet \citep{lothringer2021, chachan2023}.

We compare the retrieved atmospheric abundances of WASP-96 b with the elemental abundances of its host star, WASP-96. We denote the difference between the refractory-to-oxygen ratio of the planet and that of its host star as $\Delta\log_{10}(R/O)$. To obtain $\Delta\log_{10}(R/O)$, we first convert all the mass fractions shown to Volume Mixing Ratios (VMR) (shown in Table \ref{tab:species_combined}) using
\begin{equation}
    \text{VMR}_i = X_i \frac{\mu}{\mu_i},
\end{equation}
where $X_i$ is the mass fraction of species $i$; $\mu_i$ is the molar mass of species $i$; and $\mu$ is the atmospheric mean molar mass. We combine the abundances of sodium and potassium to obtain an estimate of the refractory abundances. Similarly, we combine the retrieved abundances of H$_2$O, CO$_2$, and CO to obtain an estimate of the oxygen abundance. Combining the posteriors of refractory and oxygen abundance that we obtain above, we calculate the refractory to oxygen abundance. Finally, we divide this refractory to the oxygen abundance with stellar abundances that we calculate in Section \ref{sec:stellar}. Due to the absence of reliable potassium absorption lines in the Magellan/Mike spectrum of WASP-96, we adopt the solar K abundance from \citep{lodders2021} as a proxy for the stellar K abundance throughout this analysis.  As a result, our estimate of $\Delta\log_{10}(R/O)$ for WASP-96 b is 1.48$^{+0.57}_{-0.62}$. Our estimate of $\Delta\log_{10}(Na/O)$ using purely the stellar abundances and planet abundances yield similar result (1.50$^{+0.57}_{-0.62}$). This suggests that WASP-96 b likely underwent rock-rich planetesimal enrichment, which typically happens in the inner disk inside the H$_2$O iceline \citep{batygin2016,boley2016}.

Theoretically, the inner disk is thought to be carbon-depleted because inherited carbonaceous solids from the interstellar medium were irreversibly destroyed early in the disk’s evolution, when high temperatures pushed the "soot line" beyond the terrestrial planet-forming region \citep{cridland2019, li2021,okamoto2024}.  The soot line marks the radial boundary in the protoplanetary disk. Inside the soot line, refractory carbon grains are thermally destroyed and converted into gas-phase species like CO and CO$_2$. Our retrieved C/O for WASP-96 b from our self-consistent chemical equilibrium retrieval ($\text{C/O}_{planet} =0.57^{+0.07}_{-0.12}$)  is consistent with the solar value, but it is low compared to its host star ($\text{C/O}_{star}= 0.92\pm0.25$). This sub-stellar C/O ratio of WASP-96 b's atmosphere suggests that the planet accreted its atmosphere from materials thermally processed through the carbon soot line. The evaporated carbon remains in gaseous form and does not recondense until much farther out in the disk, beyond the CO snowline \citep{anderson2017,krijt2020}. Consequently, unless the horizontal mixing of the disk material is high, our result suggests that the atmospheric enrichment of WASP-96 b occurred within the soot line.

Additionally, we run \texttt{GGChem} equilibrium chemistry models \citep{woitke2018} at the measured stellar C/O to assess whether WASP-96 b's carbon-depleted atmosphere results from solid enrichment in a carbon-rich disk. Our models suggest that the solid phase remains oxygen-rich even at the elevated C/O, as is the case for WASP-96. It is not until the C/O ratio exceeds $\sim$ 1.07 \citep{woitke2018} that graphite begins to condense, at which point the solid becomes carbon-rich. This implies that a sub-stellar atmospheric C/O requires enrichment by solid material that is both oxygen-rich and refractory-rich—consistent with solids that were thermally processed inside the soot line. This is consistent with our argument above.

We further attempt to infer WASP-96 b's accretion history through constraints on the planet's core mass-fraction. We use interior models from \cite{thorngren2016} with updated H/He EOS \citep{chabrier2019} and parametrization of the amount of metal in the core vs in the envelope. We use a Bayesian retrieval similar to \cite{thorngren2019a} to match our models of WASP-96 b. We input the mass from \cite{2014MNRAS.440.1982H} and the radius, equilibrium temperature, atmospheric metallcity, and age measured in this work to constrain the interior model. The posteriors are sampled with the Metropolis-Hastings MCMC algorithm. 

\begin{figure*}[t]
    \centering
    \plotone{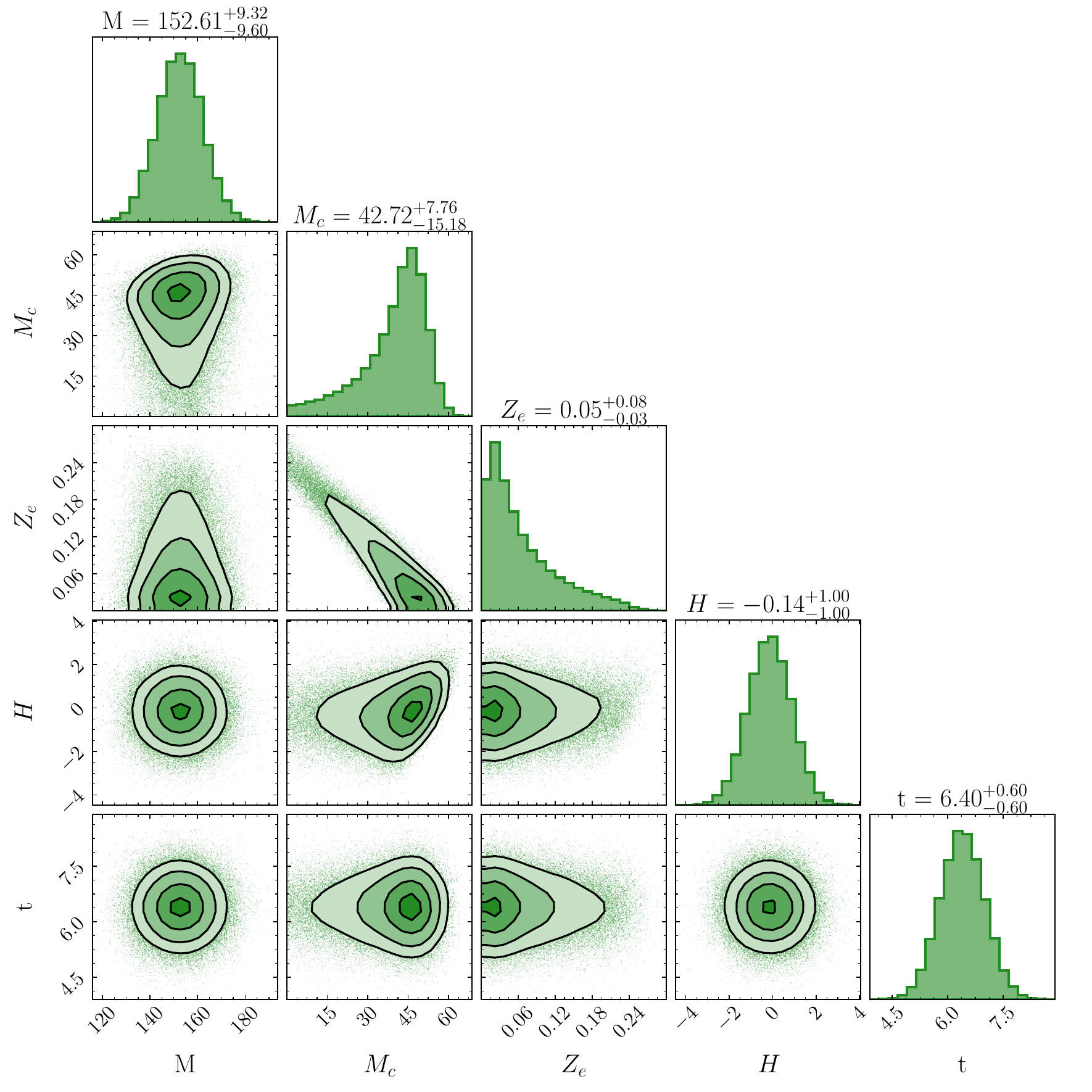}
    \caption{Corner plot of the posterior planet mass (M$_{\oplus}$), core mass (M$_{\oplus}$), envelope metallicity (unitless), heat factor (unitless), and age (Gyr) from the retrieval of the interior model.  The heat factor represents the deviation from the mean in the hot Jupiter heating distribution of \cite{thorngren2018}. Our retrieval indicates a detection of a relatively massive core (43$^{+8}_{-15}$ M$_{\oplus}$). }
    \label{fig:core}
\end{figure*}

Our interior modeling constrains the core mass at $M_c = 43^{+8}_{-15}$ M$_{\oplus}$ (Figure \ref{fig:core}). Given the planet’s relatively high density for its equilibrium temperature and the low observed atmospheric metallicity, the presence of a substantial core is therefore expected, as the low atmospheric enrichment would suggest that a significant fraction of the planet’s heavy elements is sequestered in a central core rather than mixed throughout the envelope. Forming such a massive core likely requires access to a locally enhanced reservoir of solids during the early stages of planet formation, as assembling tens of Earth masses in heavy elements is not easily achievable under standard disk conditions. This could, therefore, support the scenario in which the core of WASP-96 b formed outside of the water-ice line, where solid surface densities are enhanced by the presence of volatile ices, facilitating the rapid core accretion of tens of Earth masses of metals \citep{pollack1996}. Once the core reached a sufficient mass to trigger runaway gas accretion, the planet may have undergone disk-driven migration towards its current close-in orbit \citep{lin1996}. The planet could then have accreted a substantial portion of its atmosphere inside the carbon soot line, which could also explain the elevated $\Delta\log_{10}(R/O)$ and substellar C/O ratio.

\section{Conclusion}\label{sec:conclusion}

In this work, we have revisited the NIRISS/SOSS data of WASP-96 b with an updated calibration pipeline based on FIREFLy and combined it with existing VLT, HST, and Spitzer data to produce a panchromatic view of the planet’s atmosphere. Our reanalysis of the JWST/NIRISS SOSS observations finds no evidence for the previously reported blue optical slope and inhomogeneous clouds, which is consistent with pre-JWST results. Instead, the combined spectrum is well explained by an isothermal atmosphere at solar and substellar metallicity, C/O ratio, global gray cloud deck at moderate altitudes, and detections of Na, K, and H$_2$O. The SOSS-only data is not sufficient to confidently constrain Na and K, and the short-wavelength VLT data significantly contributes to the robustness of the detections. We attribute the updated calibration method in the \verb|jwst| pipeline as the most likely cause of the difference between our result and that reported by \cite{2023MNRAS.524..835R} and \cite{2023MNRAS.524..817T}. To isolate the effect of calibration, we re-reduced the dataset using the same reduction pipeline adopted by those studies but with the updated version of the \verb|jwst| pipeline and calibration files. This re-reduction yields a spectrum that is nearly identical to our FIREFLy result (Figure \ref{fig:reduction_latest}), demonstrating that the updated calibration alone can account for the differences in the inferred atmospheric properties. Tracking down which change made in the \verb|jwst| pipeline and calibration files led to this difference is out of the scope of this work, and requires future investigations.

We detect tentative evidence of limb asymmetry in the Na feature through the wavelength-dependent variations of the mid-transit time, which may suggest differences in sodium abundance between the morning and evening terminators. Additionally, we derive the stellar density using the JWST data and infer a precise age for WASP-96 of 6.6 $\pm$ 0.7. We also present a self-consistent elemental abundance analysis of WASP-96 using an archival Magellan/MIKE spectrum (Table \ref{tab:w96_abunds}).

Combining atmospheric and stellar characterizations, we demonstrate that we are able to speculate WASP-96 b's evolutionary history.  Our interior modeling reveals a massive core of $43^{+8}_{-15}$ M$_\oplus$, and our atmospheric retrievals suggest a super-stellar refractory-to-oxygen ratio $\Delta\log_{10}(R/O)$ and a substellar C/O ratio, although no carbon-bearing species are confidently constrained at the resolution given by Spitzer and the C/O ratio is informed solely by the assumption of chemical equilibrium. Together, these findings imply that WASP-96 b may have undergone disk-driven migration from outside the water-ice line, where high solid surface densities contribute to the buildup of the massive core, to inside the carbon soot line, where the planet accreted its refractory-rich and carbon-poor atmosphere. These combined insights provide important constraints on the formation and migration history of WASP-96 b.

The SOSS-only data is only able to provide an upper limit of CO$_2$, CO, CH$_4$, and H$_2$S, while the combined spectrum provides tighter constraints of these species. However, these species are not robustly constrained at the resolution afforded by Spitzer.  The upcoming JWST/NIRSPec G395H that covers 2.87 - 5.14 $\mu m$ will inform us more definitively of the presence of these carbon-bearing species.

\section*{ACKNOWLEDGEMENTS}

We thank the anonymous referee for their helpful comments, which significantly improved the quality of this paper.

This work is based on observations made with the NASA/ESA/CSA James Webb Space Telescope. The data were obtained from the Mikulski Archive for Space Telescopes at the Space Telescope Science Institute, which is operated by the Association of Universities for Research in Astronomy, Inc., under NASA contract NAS 5-03127 for JWST.  This work is based on observations with the NASA/ESA Hubble
Space Telescope (GO-15469, \citealt{nikolovhst}), obtained at the
Space Telescope Science Institute (STScI) operated by AURA, Inc.
This work is based on observations made with the Spitzer Space
Telescope \cite{nikolovspitzer}, which is operated by the Jet Propulsion Laboratory, California Institute of Technology, under a contract with NASA. This paper includes data gathered with the 6.5-meter Magellan Telescopes located at Las Campanas Observatory, Chile. 

Some of the data presented in this article were obtained from the Mikulski Archive for Space Telescopes (MAST) at the Space Telescope Science Institute. The specific observations analyzed can be accessed via \dataset[doi: 10.17909/xpp6-xh70]{https://doi.org/10.17909/xpp6-xh70}.

L.C.W. gratefully acknowledges support from Johns Hopkins University via a Summer Bloomberg Distinguished Professor Fellowship Award, which enabled participation in Sing’s exoplanet group and facilitated the preliminary contributions to this study. L.C.W is grateful for the fruitful discussions with Adam Langeveld, Gavin Wang, Guangwei Fu, and Kevin Schlaufman. L.C.W thanks Stefanie Sun and David Tao for their music. L.C.W thanks JHU Music Dynasty for countless therapeutic sessions through A Cappella.

\appendix

\setcounter{table}{0} 
\renewcommand{\thetable}{A\arabic{table}}
\renewcommand\theHtable{Appendix.\thetable}

\setcounter{figure}{0}
\renewcommand{\thefigure}{A\arabic{figure}}
\renewcommand\theHfigure{Appendix.\thefigure}

\section{Additional Plots}
\begin{figure*}[h!]
    \centering
    \plotone{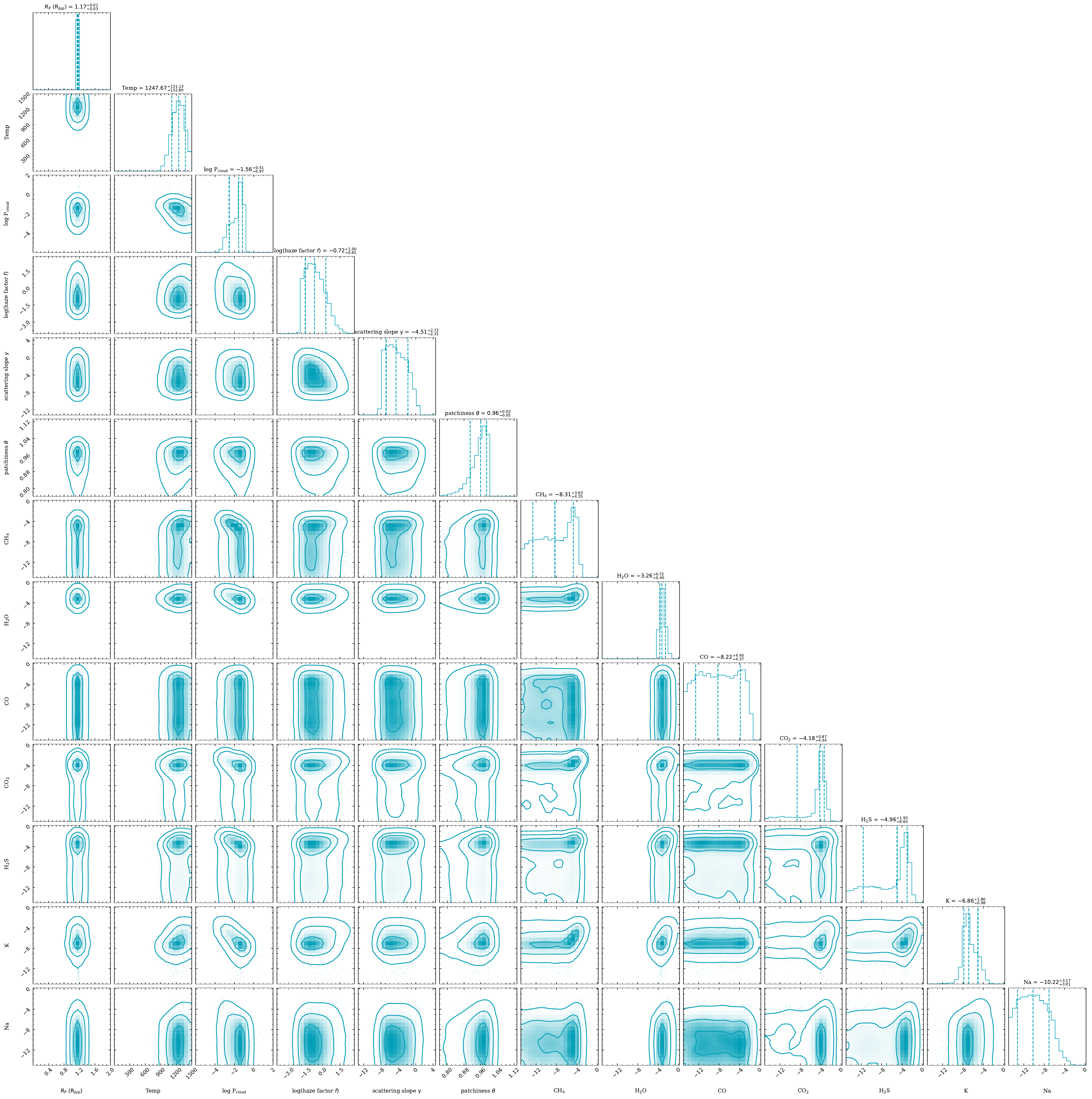}
    \caption{Corner plot for free retrieval with only JWST NIRISS/SOSS data. The retrieval is performed with petitRADTRANS with the isothermal PT-profile. The result is obtained from our hierarchical model 4 (isothermal model with inhomogeneous clouds and hazes).}
    \label{fig:soss_only_retrieval}
\end{figure*}

\begin{figure*}[h!]
    \centering
    \plotone{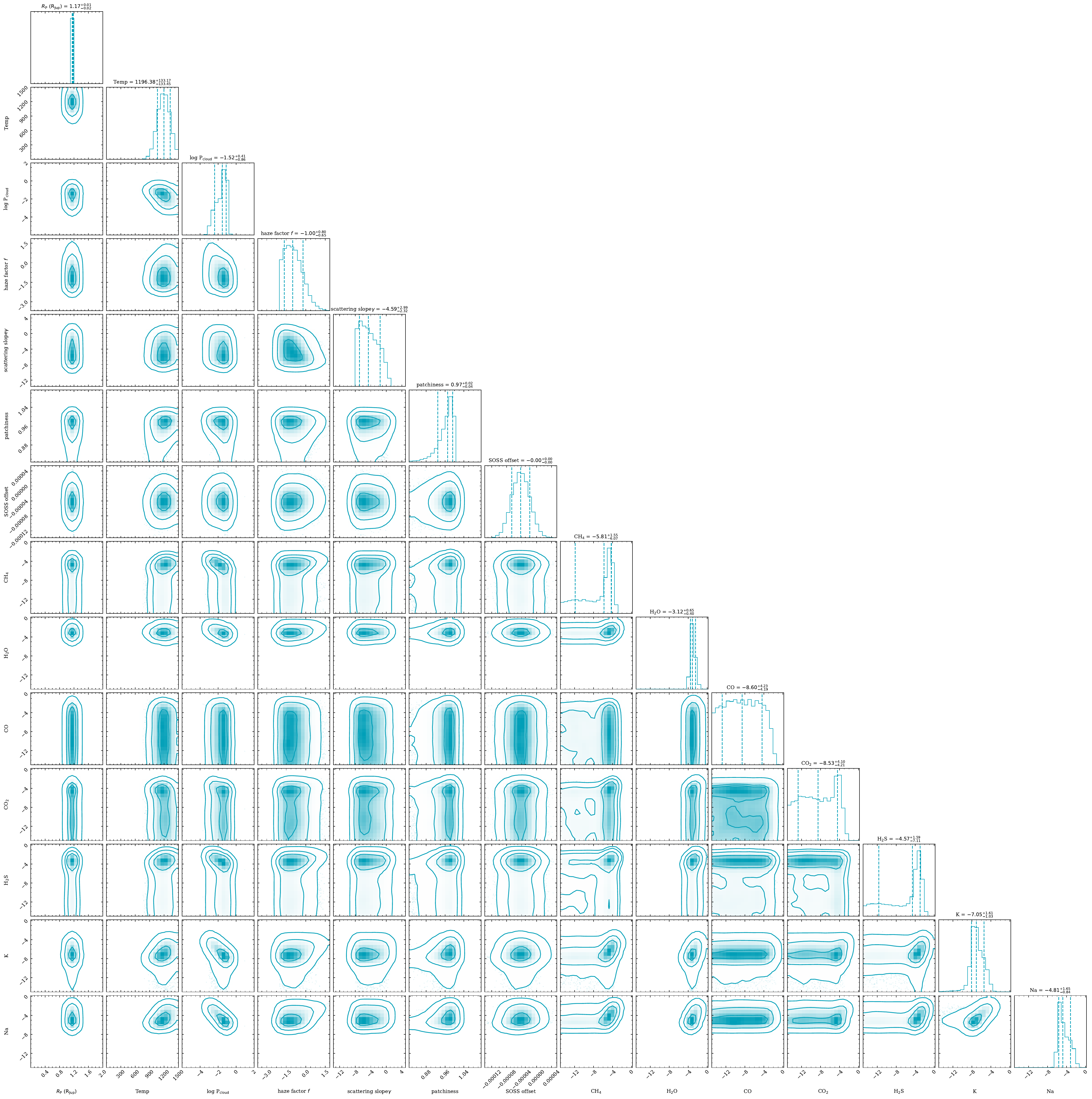}
    \caption{Similar to Figure \ref{fig:soss_only_retrieval}, but showing corner plot for free retrieval with joint JWST NIRISS/SOSS, HST, VLT, and Spitzer data.}
    \label{fig:joint_retrieval}
\end{figure*}

\begin{figure*}[h!]
    \centering
    \plotone{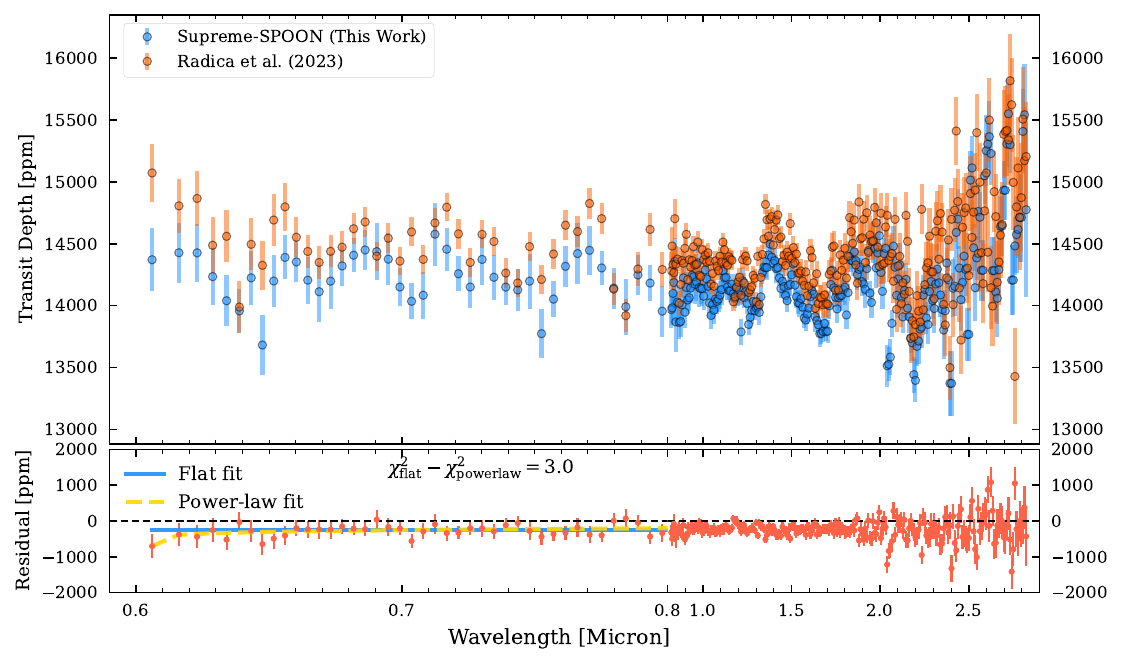}
    \caption{Same as Figure \ref{fig:vs radica} and Figure \ref{fig:reduction_latest}, but comparing the NIRISS/SOSS spectrum published by \cite{2023MNRAS.524..835R} and that reduced by Supreme-SPOON using the most up-to-date JWST calibration methods. We perform the same power-law fit vs flat line fit to the residuals as in previous figures; the power-law provides a better fit with $\Delta\chi^2$ of 3.0. The presence of a power-law slope seen at the residuals in both this figure and Figure \ref{fig:vs radica} isolates JWST calibration methods as the likely cause of such difference.}
    \label{fig:supreme_spoon_comparison}
\end{figure*}

\bibliography{main}{}
\bibliographystyle{aasjournal}


\listofchanges

\end{document}